\documentclass[aps,prc,twocolumn,showpacs,floatfix,amssymb,amsmath,10pt]{revtex4-1}

\usepackage{times,txfonts}
\usepackage{graphicx}
\usepackage{bm}
\usepackage[usenames,dvipsnames]{xcolor}
\usepackage[geometry]{ifsym}
\usepackage{xspace}

\newcommand{\dx}{\mathrm{d}}

\newcommand{\e}[1]{\mathrm{e}^{#1}}

\newcommand{\mat}[1]{\ensuremath{\begin{pmatrix} #1 \end{pmatrix}}}
\newcommand{\rpa}{\mathrm{RPA}}
\newcommand{\hf}{\mathrm{HF}}
\newcommand{\op}[1]{\hat{#1}}

\renewcommand{\H}{\op{H}}
\newcommand{\Qc}{\op{Q}^\ensuremath{\dagger}_\ensuremath{\omega}}
\newcommand{\Qd}{\op{Q}_\ensuremath{\omega}}

\newcommand{\ac}{\op{a}^\ensuremath{\dagger}}
\newcommand{\ad}{\op{a}}

\newcommand{\mlabel}[1]{\label{#1}}

\newcommand{\eMax}{\ensuremath{e_\text{max}}}
\newcommand{\lMax}{\ensuremath{l_\text{max}}}
\newcommand{\ho}{\ensuremath{\text{HO}}}

\newcommand{\lo}{\ensuremath{\text{lor}}}
\newcommand{\MeV}{\ensuremath{\text{MeV}}}
\newcommand{\keV}{\ensuremath{\text{keV}}}
\newcommand{\fm}{\ensuremath{\text{fm}}}

\newcommand{\UV}{\mbox{UCOM(VAR)$_{2b}$}\xspace}
\renewcommand{\SS}{\mbox{S-SRG$_{3b}$}\xspace}
\newcommand{\SU}{\mbox{S-UCOM(SRG)$_{3b}$}\xspace}

\newcommand{\nC}{$^\text{12}$C\xspace}
\newcommand{\nO}{$^\text{16}$O\xspace}
\newcommand{\nNe}{$^\text{20}$Ne\xspace}

\newcommand{\nSi}{$^\text{28}$Si\xspace}

\newcommand{\nS}{$^\text{32}$S\xspace}

\newcommand{\nCa}{$^\text{40}$Ca\xspace}

\definecolor{mplCr}{rgb}{1.0, 0.0, 0.0}
\definecolor{mplCg}{rgb}{0.0, 0.5, 0.0}
\definecolor{mplCb}{rgb}{0.0, 0.0, 1.0}
\definecolor{mplCc}{rgb}{0.0, 0.75, 0.75}
\definecolor{mplCm}{rgb}{0.75, 0, 0.75}
\definecolor{mplCy}{rgb}{0.75, 0.75, 0}
\definecolor{mplCk}{rgb}{0.0, 0.0, 0.0}
\definecolor{mplCw}{rgb}{1.0, 1.0, 1.0}
\definecolor{mplCgray}{gray}{0.5}

\definecolor{mplCbg}{rgb}{0.0, 1.0, 0.0}
\definecolor{mplCbm}{rgb}{1.0, 0.0, 1.0}

\definecolor{mplCb45}{rgb}{0.2203189997,0.4601782267,1}
\definecolor{mplCy75}{rgb}{0.9999305776,0.7246520567,0.225}

\definecolor{myCol}{rgb}{0.6, 0.0, 0.2}

\newcommand{\mplStu}[1][black]{\raisebox{-0.4ex}{\color{#1}\FilledSmallTriangleUp}}

\newcommand{\mplStl}[1][black]{\raisebox{-0.475ex}{\color{#1}\FilledSmallTriangleLeft}}
\newcommand{\mplStr}[1][black]{\raisebox{-0.475ex}{\color{#1}\FilledSmallTriangleRight}}

\newcommand{\mplSc}[1][black]{\raisebox{-0.5ex}{\color{#1}\FilledSmallCircle}}
\newcommand{\mplStuo}[1][black]{\raisebox{-0.4ex}{\color{#1}\textbf{\SmallTriangleUp}}}

\newcommand{\mplStlo}[1][black]{\raisebox{-0.44ex}{\rotatebox[origin=c]{90}{\color{#1}\textbf{\SmallTriangleUp}}}}
\newcommand{\mplStro}[1][black]{\raisebox{-0.52ex}{\rotatebox[origin=c]{-90}{\color{#1}\textbf{\SmallTriangleUp}}}}

\newcommand{\linethinS}[2][black]{\unitlength 1ex
  {\color{#1}
  \begin{picture}(3,1)
  \linethickness{0.3mm}
  \put(0,0.6){\line(1,0){3.0}}
  \put(0.4,0){#2}
  \end{picture}}\nolinebreak
}

\newcommand{\lineS}[2][black]{\unitlength 1ex
  {\color{#1}
  \begin{picture}(3,1)
  \linethickness{0.4mm}
  \put(0,0.6){\line(1,0){3.0}}
  \put(0.4,0){#2}
  \end{picture}}\nolinebreak
}

\newcommand{\linedashed}[1][black]{\unitlength 1ex
  {\color{#1}
  \begin{picture}(3,1)
  \linethickness{0.4mm}
  \put(0,0.6){\line(1,0){1}}
  \put(2,0.6){\line(1,0){1}}
  \end{picture}}\nolinebreak
}

\newcommand{\linedashdotted}[1][black]{\unitlength 1ex
  {\color{#1}
  \begin{picture}(3,1)
  \linethickness{0.4mm}
  \put(0,0.6){\line(1,0){1.5}}
  \put(2.5,0.6){\line(1,0){0.5}}
  \end{picture}}\nolinebreak
}

\newcommand{\linedotted}[1][black]{\unitlength 1ex
  {\color{#1}
  \begin{picture}(3,1)
  \linethickness{0.4mm}
  \put(0,0.6){\line(1,0){0.6}}
  \put(1.2,0.6){\line(1,0){0.6}}
  \put(2.4,0.6){\line(1,0){0.6}}
  \end{picture}}\nolinebreak
}

\begin{document}

\title{Collective excitations in deformed sd-shell nuclei from realistic interactions}

\author{Bastian Erler}
\email{bastian.erler@physik.tu-darmstadt.de}

\author{Robert Roth}
\email{robert.roth@physik.tu-darmstadt.de}

\affiliation{Institut f\"ur Kernphysik, Technische Universit\"at Darmstadt,
64289 Darmstadt, Germany}

\date{\today}

\begin{abstract}
\begin{description}
\item[Background] Collective excitations of nuclei and their theoretical descriptions provide an insight into the structure of nuclei. Replacing traditional phenomenological interactions with unitarily transformed realistic nucleon-nucleon interactions increases the predictive power of the theoretical calculations for exotic or deformed nuclei.
\item[Purpose] Extend the application of realistic interactions to deformed nuclei and compare the performance of different interactions, including phenomenological interactions, for collective excitations in the sd-shell.
\item[Method] Ground-state energies and charge radii of \nNe, \nSi and \nS are calculated with the Hartree-Fock method. Transition strengths and transition densities are obtained in the Random Phase Approximation with explicit angular-momentum projection.
\item[Results] Strength distributions for monopole, dipole and quadrupole excitations are analyzed and compared to experimental data. Transition densities give insight into the structure of collective excitations in deformed nuclei. 
\item[Conclusions] Unitarily transformed realistic interactions are able to describe the collective response in deformed sd-shell nuclei in good agreement with experimental data and as good or better than purely phenomenological interactions. Explicit angular momentum projection can have a significant impact on the response.
\end{description}
\end{abstract}

\pacs{21.60.Jz,24.30.Cz,27.30+t,21.30.Fe}

\maketitle

\section{Introduction}

Excited states are one of the main sources of information on the structure of atomic nuclei and are the subject of constant research in theory and experiment. Collective excitations constitute a specific class of excitations, which probe the global structure of the nucleus and allow for a geometric interpretation in terms of oscillations of intrinsic nuclear shapes. A well tested approach to describe collective excitations is the Random Phase Approximation (RPA) \cite{Rowe70}, where excited states are described by coherent particle-hole excitations starting from a mean-field-type ground state usually obtained within the Hartree-Fock (HF) approximation.

Traditionally, HF and RPA calculations are being performed with phenomenological interactions.
The form of such interactions is guided by symmetry considerations and computational simplicity, and their parameters are typically fitted to binding energies and radii of a series of nuclei within the chosen approximation for treating the many-body problem, e.g., the HF approximation. Popular phenomenological interactions are the various Skyrme forces, e.g. \cite{Skyrme58,Chabanat98,Tondeur00}, and the Gogny D1S interaction \cite{DG80}. These interactions allow for efficient HF calculations of nuclear binding energies and other ground-state properties and typically yield good agreement with experiment. However, when applying these interactions in another many-body scheme, like RPA, or to other observables, like the collective response, their performance might deteriorate.

An alternative approach, which we follow in this work, uses unitarily transformed realistic nucleon-nucleon interactions, like the phenomenological Argonne V18 (AV18) \cite{Wiringa95}, CD Bonn \cite{Machleidt01} or Nijmegen \cite{Stoks94} interactions, or interactions derived from chiral effective field theory (EFT) \cite{Epelbaum02,Entem03,Epelbaum05,Machleidt10}.
These interactions are not tuned to specific properties of finite nuclei and are not determined within a specific approximation scheme, but rather are fit to two-nucleon phase-shifts and deuteron properties in exact calculations. Thus, these interactions are universal and can be employed in different many-body approaches to describe different states and observables on equal footing. The unitary transformations help to improve the convergence of the many-body calculations with respect to the many-body model space. At the same time, the quality of simple approximations, like the HF approximation, is improved \cite{Roth10,Roth06}. The interactions used in this work are based on the AV18 potential, transformed either with the Unitary Correlation Operator Method (UCOM) or the Similarity Renormalization Group (SRG) method \cite{Roth10,Roth08}.

During the past decade, routine calculations for deformed nuclei within the HF and RPA framework have become possible.
So far, these calculations have only been carried out with phenomenological interactions \cite{Peru08,Arteaga08,Losa10}.
This work is the first application of the HF-RPA treatment with unitarily transformed realistic interactions to deformed nuclei.
The extension to intrinsically deformed nuclei opens a new domain of application away from semi-magic nuclei and allows for a detailed study of the impact of deformation on collective modes. Due to the symmetry breaking in the mean-field density, the HF state is no longer an eigenstate of the angular-momentum operator and the proper symmetry has to be restored by an explicit angular-momentum projection.

In this work, we study nuclei in the sd-shell. These nuclei already exhibit strong deformations with only a small number of nucleons. We focus on three even-even self-conjugate nuclei, the prolate nuclei \nNe and \nS, and the oblate nucleus \nSi.
\nNe and \nSi show a purely axial deformation, while \nS also has a small triaxial component. The deformation in these nuclei is driven by $\alpha$-cluster correlations in the ground states, which makes them a particularly interesting candidate for the study of deformation effects on the collective response. At the same time the presence of $\alpha$-clustering suppresses the pairing correlations in these open-shell nuclei, so that a simple HF approach without explicit pairing is applicable. For these first calculations we restrict ourselves to axial deformations, which simplifies the explicit angular-momentum projection significantly.

\section{Hartree Fock method}\label{hf}
\subsection{Formalism}

To obtain a nuclear ground-state, the starting point for the RPA, we employ the HF method.
We use $ls$-coupled spherical harmonic oscillator (HO) states as the computational basis.
An HO state is fully determined by the quantum numbers $n$, $l$, $j$, $m_j$ and $m_t$.
In a spherical HF implementation, only states with different $n$ but equal $l$, $j$ and $m_j$ can mix.
Deformed HF states can be obtained, if the HF single-particle states are also allowed to contain contributions of different total and orbital angular-momentum $j$, $m_j$ and $l$
\begin{align}
    |\ensuremath{\alpha}\,m_t\ensuremath{\rangle}=\sum_{a} C^{\ensuremath{\alpha}\,m_t}_a |a\,m_t\ensuremath{\rangle} \;.
\end{align}
The HO quantum numbers $n$, $l$, $j$ and $m_j$ are combined in the index $a$.
The HF state index $\ensuremath{\alpha}$ covers the same range as the HO index $a$, but does not have the same physical meaning.
If a nucleus attains an axially symmetric deformation, the HF single-particle states are a superposition of HO states with different $n$, $l$, $j$. The single-particle projection quantum numbers $m_j$ remain good quantum numbers, and their sum in the HF state, denoted by $K$, defines the angular-momentum projection onto the symmetry axis of the intrinsic frame, which is the only remaining good quantum number in the intrinsic frame (except for isospin).

To obtain quantities in the lab-frame, where the ground state is an eigenstate of the total angular-momentum operator $\op{J}^{2}$, angular-momentum projection has to be employed \cite{Loewdin55-1,Brink66,Ring80}.
The angular-momentum projected energy of an axially symmetric nucleus is
\begin{align}
    E^{J} = \frac{\ensuremath{\langle}\hf|\op{H} \, \op{P}^J_{KK}|\hf\ensuremath{\rangle}}{\ensuremath{\langle}\hf|\op{P}^J_{KK}|\hf\ensuremath{\rangle}} \;.
\end{align}
The projection operator for axial-symmetry is given by
\begin{align}\mlabel{proj:defp-as}
    \op P^J_{MK} &= \frac{2\,J+1}{2} \, \int_{-1}^1 \, d^J_{MK}(\ensuremath{\beta}) \, \e{i\ensuremath{\beta}\op{J}_y} \, \dx(\cos\ensuremath{\beta}) \;,
\end{align}
where $d^J_{MK}(\ensuremath{\beta})$ denotes the reduced Wigner-Functions.

The ground state is obtained by minimizing the projected ground-state energy in a so called \textit{variation-after-projection} approach.
This procedure is approximated by carrying out a number of constrained HF calculations with the modified Hamiltonian
\begin{align}\mlabel{proj:hamil}
    \H' = \H - \ensuremath{\lambda}\,\op Q \;.
\end{align}
Among these solutions, the one with the lowest projected ground-state energy is selected.
The quadrupole operator $\op Q$ is a natural choice for the constraint, as this is the dominant collective degree of freedom for axially deformed nuclei.
We refer to this treatment as \textit{approximate variation-after-projection}.

\subsection{Calculation details}
The HO basis used for our calculations is truncated with respect to the principal oscillator quantum number $2\,n+l=e\leq\eMax$, with an additional truncation for the orbital angular-momentum $l\leq\lMax$.
Unless stated otherwise, we use $\eMax=14$ and $\lMax=10$.
The ground-state energies for $\eMax=14$ are converged to within less than $50\:\keV$.
The optimal harmonic oscillator lengths $a_\ho$ are determined by a minimization of the ground-state energy over a set of discrete oscillator lengths, the values used throughout this work are summarized in Tab. \ref{aHOtab}.

We use a total of four interactions in this work.
Three are based on unitary transformations of the AV18 potential.
\UV is a pure two-body interaction, transformed with the UCOM, where the correlation operators are determined from a variational approach.
It was first published in 2005 \cite{Roth05} and has since been used in a number of calculations \cite{Roth06,Paar06,Papakonstantinou07,Papakonstantinou10,Papakonstantinou11} and is also described in \cite{Roth10}.
Since the \UV interaction does not reproduce the correct charge radii (cf. section \ref{s:Results}), other interactions have been developed, which go beyond a pure two-body interaction, e.g., the \SS and \SU interactions introduced in \cite{Gunther10}.
The \SS interaction is transformed via an SRG flow-evolution.
For the \SU interaction, the solution of an SRG flow-evolution is used to determine the correlation operators for the UCOM transformation.
In both interactions, the unitary transformation only acts on partial waves containing relative S waves, i.e., the $^1S_0$ and
the coupled $^3S_1$-$^3D_1$ partial waves.
To include effects of three-body forces, these interactions also contain a simple phenomenological three-body contact force $\op V_3 = C_3\,\ensuremath{\delta}(\op r_1-\op r_2)\,\ensuremath{\delta}(\op r_2-\op r_3)$.
The strength $C_3$ is adjusted to reproduce the charge radius systematics for doubly-magic nuclei from $^4\text{He}$ to $^{208}\text{Pb}$, details can be found in Ref.~\cite{Gunther10}. First applications of these interactions in the RPA framework for collective excitations in spherical nuclei are discussed in Ref.~\cite{Gunther14}.
The contact interaction is used as a computationally efficient substitute for realistic three-body forces.
In mean-field calculations, contact interactions of arbitrary particle rank can be reduced to lower-order interactions which only depend on integer powers of the ground-state density.
The inclusion of realistic three-body interactions, e.g. the ones from chiral EFT, is still extremely challenging for deformed nuclei.
As an example of a popular phenomenological interaction, we also included the Gogny D1S interaction \cite{DG80} into our studies.

\begin{table}
\begin{tabular}{lcccc}
\toprule
 & & \makebox[0pt][l]{interaction} & & \\
\cline{2-5}
 & \makebox[1.7cm][l]{\UV} & & \makebox[1.7cm][c]{\SU} & \\
 nucleus & & \makebox[1.7cm][c]{\SS} & & \makebox[1.7cm][r]{Gogny D1S} \\
\colrule
\nC  & $1.6$ & $1.6$ & $1.6$ & $1.5$ \\
\nO  & $1.6$ & $1.6$ & $1.6$ & $1.5$ \\
\nNe & $1.6$ & $1.6$ & $1.6$ & $1.5$ \\
\nSi & $1.6$ & $1.7$ & $1.7$ & $1.5$ \\
\nS  & $1.6$ & $1.7$ & $1.8$ & $1.6$ \\
\nCa & $1.6$ & $1.8$ & $1.8$ & $1.6$ \\
\botrule
\end{tabular}
\caption{Optimal oscillator lengths ($a_\ho$) for the various nuclei and interactions (in \fm).}
\label{aHOtab}
\end{table}

\subsection{Results}\label{s:Results}

\begin{figure*}[th]
    \includegraphics{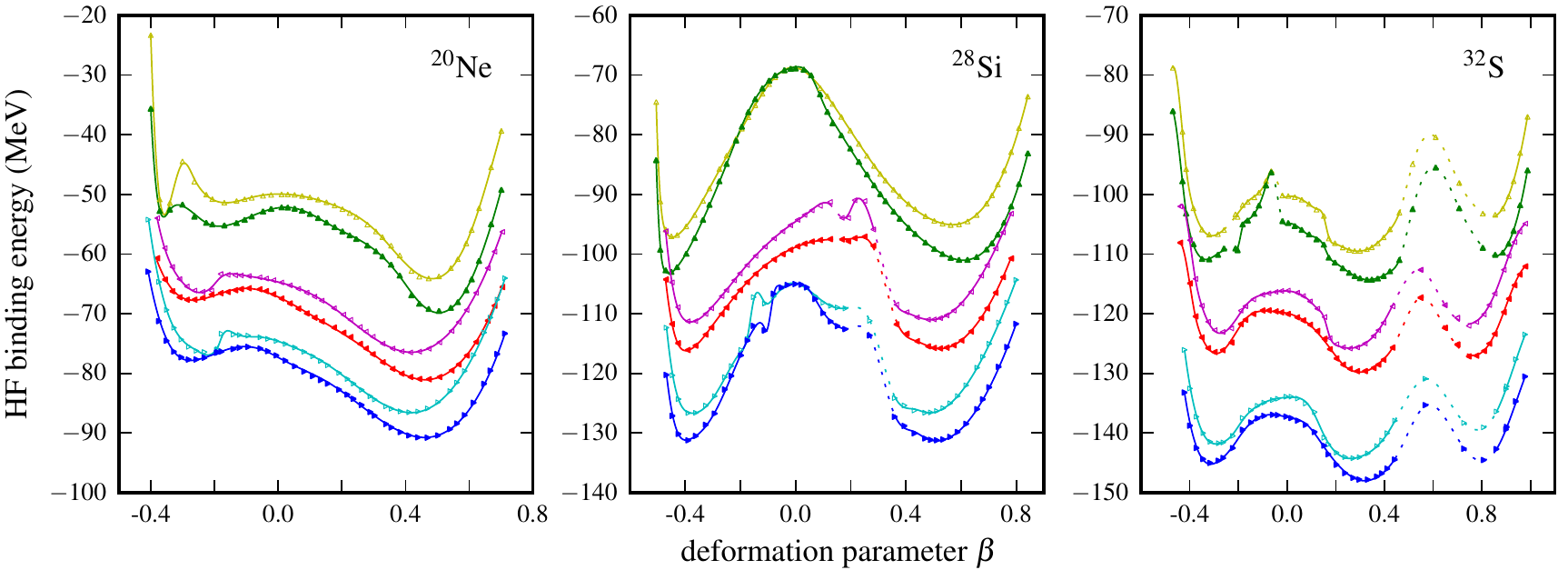}
    \caption{(color online) Intrinsic and angular-momentum projected ground-state energy as a function of axial deformation for different interactions. Markers denote actual data points, lines are drawn with a spline approximation, dashed lines indicate insufficient data points. The three interactions are: \UV (\linethinS[mplCy]{\mplStuo[mplCy]}) and (\linethinS[mplCg]{\mplStu[mplCg]}), \SS (\linethinS[mplCm]{\mplStlo[mplCm]}) and (\linethinS[mplCr]{\mplStl[mplCr]}), \SU (\linethinS[mplCc]{\mplStro[mplCc]}) and (\linethinS[mplCb]{\mplStr[mplCb]}). This plot uses a basis truncation of $\eMax=\lMax=10$.}
    \label{Ebeta}
\end{figure*}

\begin{figure*}[th]
    \centering
    \includegraphics{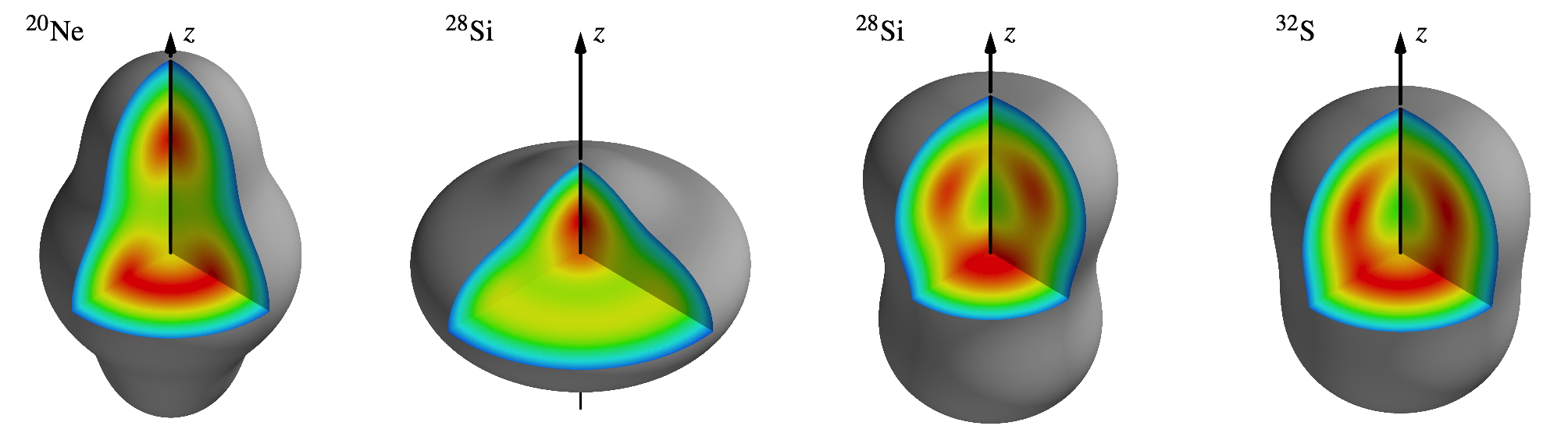}
    \caption{(color online) Density distribution for energy minima (oblate and prolate for \nSi) obtained with the \SU interaction. The isosurface is located at $40\%$ of maximum density and the maximum densities are $0.169\,\fm^{-3}$ for \nNe, $0.190\,\fm^{-3}$ for oblate \nSi, $0.162\,\fm^{-3}$ for prolate \nSi and $0.161\,\fm^{-3}$ for \nS.}
    \label{rhoGS}
\end{figure*}

\begin{figure}[ht]
    \centering
    \includegraphics{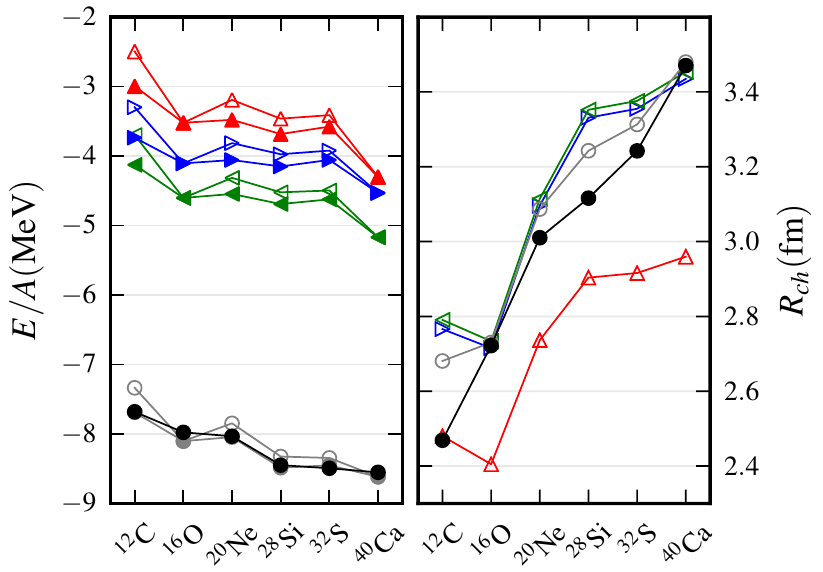}
    \caption{(color online) Systematics of the projected ground-state energy per nucleon and the nuclear charge-radius for the different interactions used in this study: \UV (\linethinS[mplCr]{\mplStu[mplCr]}), \SS (\linethinS[mplCb]{\mplStr[mplCb]}), \SU (\linethinS[mplCg]{\mplStl[mplCg]}), Gogny D1S (\linethinS[mplCgray]{\mplSc[mplCgray]}), experiment (\linethinS[mplCk]{\mplSc[mplCk]}). Open symbols intrinsic, filled symbols projected values.}
    \label{Esys}
\end{figure}

We start by discussing the HF results for the ground states of \nNe, \nSi and \nS with and without angular-momentum projection. 
Figure \ref{Ebeta} shows the dependence of the ground-state energy on the deformation parameter.
Here, we used a truncation of $\eMax=10$ to reduce the computational cost. 
At this level, the ground-state energy is converged within $1\:\MeV$, which is sufficient for the presentation in Fig. \ref{Ebeta}.
The ground-state energy of \nNe varies continuously with the deformation, with a minimum at a strong prolate deformation.
For \nSi and \nS, we see some irregularities in the curves, which are an artifact of the constrained minimization procedure. From a practical point of view this does not pose a problem, as long as the discontinuities are sufficiently far away from the energy minimum, which is always the case.
\nSi shows two practically degenerate minima, one for oblate and one for prolate deformation.
For the pure two-body interaction \UV and the Gogny D1S interaction \cite{Peru08}, the oblate minimum has a slightly lower energy and both minima are connected through triaxial deformation.
However, this is not the case for the \SS and \SU interactions.
Since the oblate shape of \nSi is well established, we focus on this solution for our RPA calculations.
\nS shows three local minima, one at oblate deformation, one at a moderate prolate deformation and one at a strong prolate deformation.
For all interactions, the minimum with the moderate prolate deformation is the absolute minimum and it is well separated from the other minima. It should be noted that the general dependence of the ground-state energy on the deformation is very robust under changes of the underlying interaction. 

The intrinsic density distributions for the ground states of the relevant minima are shown in Fig. \ref{rhoGS}. Obviously, the intrinsic shapes are more complicated than the simple ellipsoids corresponding to pure quadrupole deformations. The interactions induce $\alpha$-cluster correlations, which generate deformations of higher multipole orders as can be seen in the intrinsic densities.  

Figure \ref{Esys} shows the systematics of the ground-state energy per nucleon and the nuclear radius.
The most prominent feature of the ground-state energy per nucleon is the difference of about $4\text{--}5\:\MeV$ between the measured energies and the values obtained with realistic interactions.
This shift is due to correlations that are not described by a single Slater determinant and cannot be recovered by angular momentum projection. Since the unitary transformations only account for the short-range correlations, these missing correlations are driven by intermediate-range contributions in the interaction.
We have shown in several previous publications that these missing correlations can be described by many-body perturbation theory and that the inclusion of low-order perturbative corrections to the energy leads to a good systematic agreement of the ground-state energies with experiment \cite{Gunther10,Roth10,Roth05}. We have also shown that the RPA does describe these ground-state correlations very well and that the inclusion of the RPA correlation energy (ring summation) leads to a good agreement with the experimental ground-state energies for closed-shell nuclei \cite{Barbieri06}. The contributions to the correlation energy resulting from long-range correlations related to deformation are recovered by the angular-momentum projection and are significantly smaller than the intermediate-range correlations.

In case of the charge-radii, the missing correlations only play a minor role and the difference between the intrinsic and projected radii is negligible.
For spherical nuclei, the two- plus three-body interactions \SS and \SU are in good agreement with experiment (see also \cite{Gunther10}).
While the radii of \nNe and \nS are still described rather well, the results for \nSi are about 10\% above the measured radii.

Figure \ref{Esys} also contains data obtained with the Gogny D1S interaction.
Since the interaction is fitted to binding energies, these are reproduced very accurately.
The radii of deformed nuclei are comparable to those obtained with the \SS and \SU interactions.

\section{Random phase approximation for deformed nuclei}

\subsection{Formalism}

In the standard RPA, excitations are described as one-particle one-hole and one-hole one-particle excitations.
The excitation operator $\Qc$ is given by
\begin{equation} \mlabel{rpa:phonon}
    \Qc = \sum_{mi} X^\ensuremath{\omega}_{ma}\,\ac_m\ad_a - \sum_{ma} Y^\ensuremath{\omega}_{ma}\,\ac_a\ad_m \;,
\end{equation}
where indices starting with $m$ denote states above the Fermi level and those starting with $a$ denote states below. The RPA ground-state is defined by the relation $\Qd |\rpa\ensuremath{\rangle} = 0$ and excited states are given by $|\ensuremath{\omega}\ensuremath{\rangle}=\Qc|\rpa\ensuremath{\rangle}$.

The summation in \eqref{rpa:phonon} runs over all possible particle-hole (ph) pairs defined with respect to the HF ground state. In the case of deformed ground-states, the number of ph-pairs cannot be reduced by angular momentum coupling rules, as $j$ ceases to be a good quantum number. However, for axial deformations, the number of ph-pairs can still be reduced by considering the projection quantum number $m_j$ and parity.
In a spherically symmetric basis spanning 15 major HO shells, calculating electric monopole excitations requires $38$ ph pairs for \nO and $72$ for \nCa.
If the basis is extended to allow axially symmetric deformations, these numbers increase to $780$ and $1968$, respectively.

The amplitudes $X^\ensuremath{\omega}_{ma}$ and $Y^\ensuremath{\omega}_{ma}$ are obtained by the equations-of-motion method and the quasi-boson approximation \cite{Ring80}, which result in the RPA matrix equation
\begin{equation} \mlabel{rpa:rpa-eq}
    \mat{A &  B \\ B^* & A^*} \mat{X^\ensuremath{\omega} \\ Y^\ensuremath{\omega}} = E_\ensuremath{\omega} \mat{1 & 0 \\ 0 & -1} \mat{X^\ensuremath{\omega} \\ Y^\ensuremath{\omega} } \;,
\end{equation}
with
\begin{equation}\mlabel{rpa:2bAB}
\begin{split}
    A_{manb} &= \ensuremath{\langle}\hf| [\ac_a\ad_m,[\H,\ac_n\ad_b]] |\hf\ensuremath{\rangle} \\
             &= (\ensuremath{\varepsilon}_m-\ensuremath{\varepsilon}_a)\ensuremath{\delta}_{mn}\,\ensuremath{\delta}_{ab} + \ensuremath{\langle}m,b|\H|a,n\ensuremath{\rangle} \;,\\
    B_{manb} &= -\ensuremath{\langle}\hf| [\ac_a\ad_m,[\H,\ac_b\ad_n]] |\hf\ensuremath{\rangle} \\
             &= \ensuremath{\langle}m,n|\H|a,b\ensuremath{\rangle} \;.
\end{split}
\end{equation}
Here, a Hamiltonian with only one- and two-body terms is assumed.
If the Hamiltonian also includes a three-body interaction $\op V_3$, it has to be separated and the following terms have to be added to Eq. \eqref{rpa:2bAB}
\begin{equation}\mlabel{rpa:3bAB}
\begin{split}
    A^{(3)}_{manb} &= \sum_{k} \ensuremath{\langle}m,b,k|\op V_3|a,n,k\ensuremath{\rangle} \;,\\
    B^{(3)}_{manb} &= \sum_{k} \ensuremath{\langle}m,n,k|\op V_3|a,b,k\ensuremath{\rangle} \;.
\end{split}
\end{equation}

\subsection{Projected transition matrix element}

The reduced transition probability for an electric transition operator $\op T_{\ensuremath{\lambda}\ensuremath{\mu}}$ from an initial state $\ensuremath{|\Phi_0\rangle}$ to the final state $\ensuremath{|\Phi_{\omega}\rangle}$ is defined as
\begin{equation}
\begin{split}
    B(\text{E}\ensuremath{\lambda},J_\ensuremath{0}\rightarrow J_\ensuremath{\omega})=\frac{1}{2\,J_0+1}|(\ensuremath{\Phi_0}\|\op T_\ensuremath{\lambda}\|\ensuremath{\Phi_{\omega}})|^2\\ =\frac{1}{2\,J_0+1}\sum_\ensuremath{\mu} |\ensuremath{\langle}\ensuremath{\Phi_0}|\op T_{\ensuremath{\lambda}\ensuremath{\mu}}|\ensuremath{\Phi_{\omega}}\ensuremath{\rangle}|^2 \;.
\end{split}
\end{equation}
We use the shorthand notation $B(E\ensuremath{\lambda})$ for $J_\ensuremath{0}=0$ and $J_\ensuremath{\omega}=\ensuremath{\lambda}$. 
Intrinsic transition amplitudes between the RPA ground-state and an excited state are calculated with
\begin{equation}\mlabel{rpa:rpa-int}
\begin{split}
    \ensuremath{\langle}\rpa|\op{T}_{\ensuremath{\lambda}\ensuremath{\mu}}\Qc|\rpa\ensuremath{\rangle}&\approx\ensuremath{\langle}\hf|[\op{T}_{\ensuremath{\lambda}\ensuremath{\mu}},\Qc]|\hf\ensuremath{\rangle}\\
    &= \sum_{ma} \big( X^\ensuremath{\omega}_{ma}\, \ensuremath{\langle}\hf|\op{T}_{\ensuremath{\lambda}\ensuremath{\mu}}\,\ac_m\ad_a|\hf\ensuremath{\rangle} \\
    &\phantom{= \sum_{ma} \big(} + Y^\ensuremath{\omega}_{ma}\, \ensuremath{\langle}\hf|\ac_a\ad_m\,\op{T}_{\ensuremath{\lambda}\ensuremath{\mu}}|\hf\ensuremath{\rangle} \big)\\
    &= \sum_{ma} \left( X^\ensuremath{\omega}_{ma}\, \ensuremath{\langle}a|\op{T}_{\ensuremath{\lambda}\ensuremath{\mu}}|m\ensuremath{\rangle} + Y^\ensuremath{\omega}_{ma}\, \ensuremath{\langle}m|\op{T}_{\ensuremath{\lambda}\ensuremath{\mu}}|a\ensuremath{\rangle} \right) \;.
\end{split}
\end{equation}
Transition amplitudes between angular-momentum projected RPA states, denoted by their total angular-momentum values $J_0$ and $J_\ensuremath{\omega}$, are given by the following formula \cite{Ring80} 
\begin{equation}
\begin{split}
    (J_0\|\op{T}_\ensuremath{\lambda}\|J_\ensuremath{\omega})&=(2\,J_0+1) \, N_0 \, N_\ensuremath{\omega} \,  \sum_{\substack{K_0 K_\ensuremath{\omega}\\\ensuremath{\mu}}} \,g^{(0)}_{K_0}\,g^{(\ensuremath{\omega})}_{K_\ensuremath{\omega}}\\
    &\phantom{=} \times (-1)^{\ensuremath{\lambda}+J_\ensuremath{\omega}+K_0}  \mat{J_0 & J_\ensuremath{\omega} & \ensuremath{\lambda} \\ -K_0 & K_0-\ensuremath{\mu} & \ensuremath{\mu}}\\
    &\phantom{=} \times \ensuremath{\langle}\rpa|\op{T}_{\ensuremath{\lambda}\ensuremath{\mu}}\, \op{P}^{J_\ensuremath{\omega}}_{K_0-\ensuremath{\mu}, K_\ensuremath{\omega}}\,\Qc|\rpa\ensuremath{\rangle} \;,
\end{split}
\end{equation}
with the normalization factors $N_0$ and $N_\ensuremath{\omega}$.
Similar to Eq. \eqref{rpa:rpa-int}, we express this in terms of the HF ground-state and the $X$- and $Y$-amplitudes
\begin{equation} \mlabel{rpa:proj-as}
\begin{split}
    (J_0\|\op{T}_\ensuremath{\lambda}\|J_\ensuremath{\omega}) &\approx (2\,J_0+1) \, N_0 \, N_\ensuremath{\omega} \,  (-1)^{J_0-K_\hf} \\
    &\phantom{=} \times \sum_{\substack{ma\\\ensuremath{\mu}}} \left( X^\ensuremath{\omega}_{ma} + (-1)^{K_{ma}}Y^\ensuremath{\omega}_{ma}\right)\mat{J_0 & \ensuremath{\lambda} & J_\ensuremath{\omega} \\ -K_\hf & \ensuremath{\mu} & K_\hf-\ensuremath{\mu}} \\
    &\phantom{=} \times \ensuremath{\langle}\hf|\op{T}_{\ensuremath{\lambda}\ensuremath{\mu}}\, \op{P}^{J_\ensuremath{\omega}}_{K_\hf-\ensuremath{\mu}, K_\hf+K_{ma}}\ac_m\ad_a|\hf\ensuremath{\rangle} \;.
\end{split}
\end{equation}
Notation and further details are discussed in Appendix \ref{appSTR}.
We like to point out that Eq. \eqref{rpa:proj-as} is used directly in the calculations.
We do not use any further approximations for either the overlaps or the integration involved in the angular momentum projection (the numerical integration is performed using 2048 points).

\subsection{Transition operators}

We use the standard form of the electric transition operators in the long wavelength limit given by \cite{Ring80} 
\begin{equation} \mlabel{rpa:T}
    \hat T_{\ensuremath{\lambda}\ensuremath{\mu}} = \sum^A_{i} e_i\,\op r^\ensuremath{\lambda}_i\,Y_{\ensuremath{\lambda}\ensuremath{\mu}}(\op{\ensuremath{\Omega}}_i) \;.
\end{equation}
As ususal, the electric transitions are decomposed into a sum of an isoscalar and an isovector part
\begin{align}
    \hat T_{\ensuremath{\lambda}\ensuremath{\mu}} &= \frac{1}{2}\left( \hat T_{\ensuremath{\lambda}\ensuremath{\mu}}^\text{IS} + \hat T_{\ensuremath{\lambda}\ensuremath{\mu}}^\text{IV} \right) \\
    \hat T_{\ensuremath{\lambda}\ensuremath{\mu}}^\text{IS} &= e \sum^Z_i \op r^\ensuremath{\lambda}_i\,Y_{\ensuremath{\lambda}\ensuremath{\mu}}(\op{\ensuremath{\Omega}}_i) + e \sum^N_i \op r^\ensuremath{\lambda}_i\,Y_{\ensuremath{\lambda}\ensuremath{\mu}}(\op{\ensuremath{\Omega}}_i) \\
    \hat T_{\ensuremath{\lambda}\ensuremath{\mu}}^\text{IV} &= e \sum^Z_i \op r^\ensuremath{\lambda}_i\,Y_{\ensuremath{\lambda}\ensuremath{\mu}}(\op{\ensuremath{\Omega}}_i) - e \sum^N_i \op r^\ensuremath{\lambda}_i\,Y_{\ensuremath{\lambda}\ensuremath{\mu}}(\op{\ensuremath{\Omega}}_i) \;.
\end{align}

For the electric monopole transitions the generic first-order transition operator \eqref{rpa:T} is a constant and thus cannot induce transitions. Instead the second-order term is generally used
\begin{equation} \mlabel{rpa:TE0}
    \hat T_{00} = \sum^A_i e_i\,\op r^2_i\,Y_{00}(\op{\ensuremath{\Omega}}_i) \;.
\end{equation}
Since the electric dipole operator is potentially contaminated by spurious center-of-mass contributions, corrected transition operators are used \cite{VanGiai81,Harakeh01}
\begin{align} \mlabel{rpa:TE1S}
    \hat T_{1\ensuremath{\mu}}^\text{IS} &= e \sum_i^A \left( \op r_i^3 - \tfrac{5}{3} R_{\text{ms}}\,\op r_i\right) Y_{1\ensuremath{\mu}}(\op{\ensuremath{\Omega}}_i) \;, \\ \mlabel{rpa:TE1V}
    \hat T_{1\ensuremath{\mu}}^\text{IV} &= e \frac{N}{A} \sum_i^Z \op r_i\,Y_{1\ensuremath{\mu}}(\op{\ensuremath{\Omega}}_i) - e \frac{Z}{A} \sum_i^N \op r_i\,Y_{1\ensuremath{\mu}}(\op{\ensuremath{\Omega}}_i) \;,
\end{align}
with the mean-square radius $R_{\text{ms}}$ of the nucleus.

In principle, the unitary transformation used for the interactions also has to be applied to the transition operators, but considering missing higher-order correlations in the RPA and the long-range and low-momentum character of $r^\ensuremath{\lambda}$, it is justified to neglect this transformation.
For the \UV interaction, it was shown that the correction due to transformed transition operators is indeed small and not relevant in most cases \cite{Paar06}.

\section{RPA results}
\subsection{Convergence and sensitivity}

\begin{figure}[ht]
    \centering
    \includegraphics{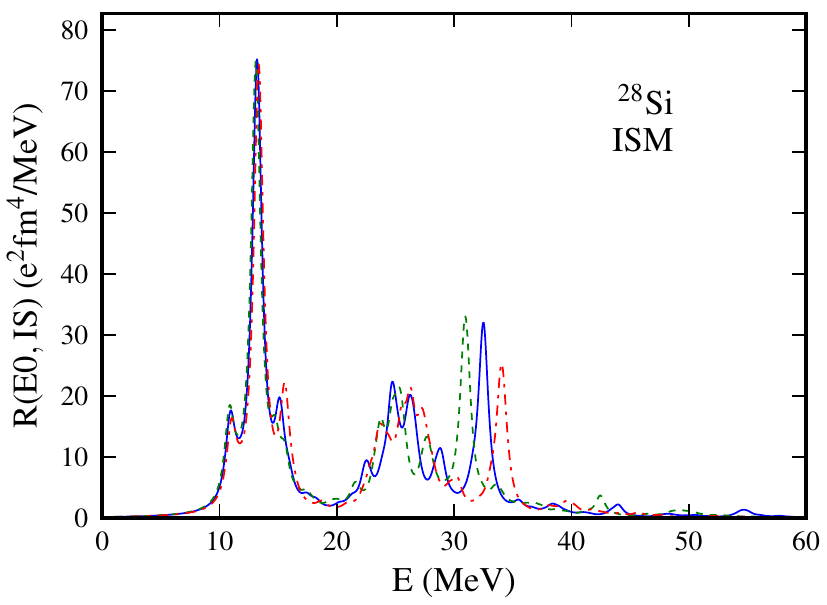}
    \caption{(color online) Convergence of the response function for the ISM mode in \nSi with respect to basis truncation and harmonic oscillator length: $\eMax=14$, $a_\ho=1.7\,\fm$  (\lineS[mplCb]{}); $\eMax=14$, $a_\ho=1.8\,\fm$  (\linedashed[mplCg]{}); $\eMax=12$, $a_\ho=1.7\,\fm$  (\linedashdotted[mplCr]{}), \SU interaction.}
    \label{convEaHO}
\end{figure}

\begin{figure}[ht]
    \centering
    \includegraphics{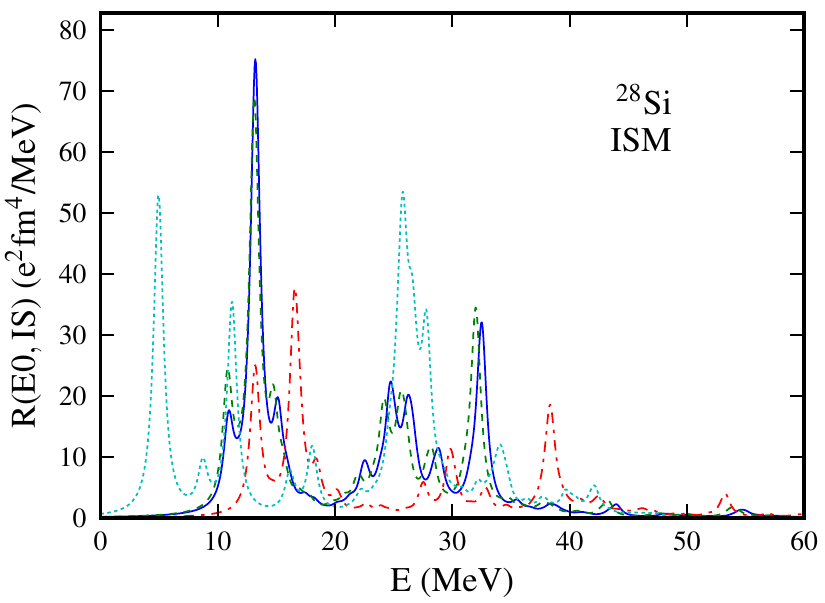}
    \caption{(color online) Response functions for the ISM mode in \nSi for different interactions: \SU (\lineS[mplCb]{}) , \SS (\linedashed[mplCg]{}), \UV (\linedashdotted[mplCr]{}) and Gogny D1S (\linedotted[mplCc]{}).}
    \label{convME}
\end{figure}

\begin{figure*}[t]
    \centering
    \includegraphics{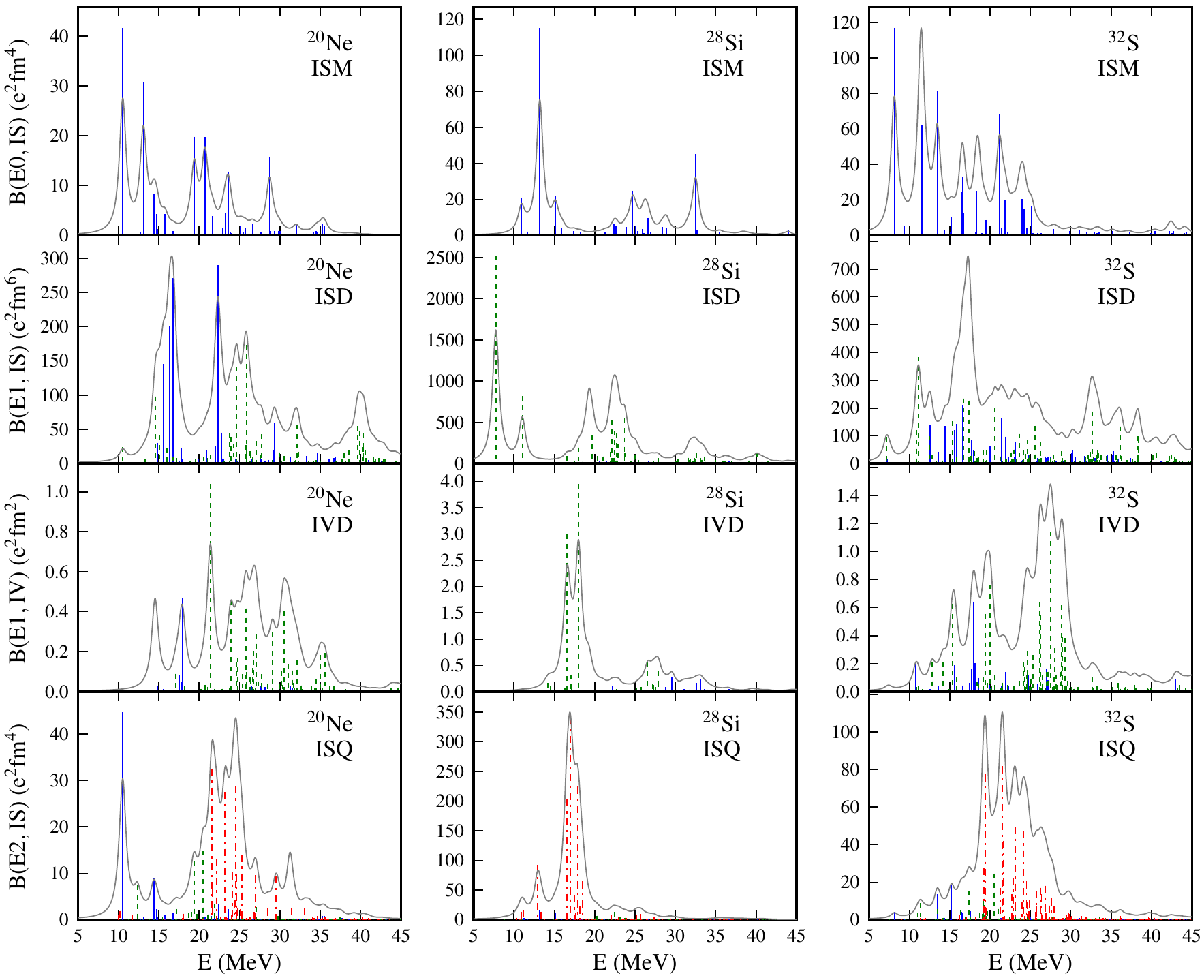}
    \caption{(color online) Response functions for the ISM, ISD, IVD, and ISQ excitations (top to bottom) for \nNe, \nSi, and \nS (left to right) obtained in angular-momentum projected RPA calculations with the \SU interaction. The $K$-components are: $K=0$ (\lineS[mplCb]{}), $K=\ensuremath{\pm}1$ (\linedashed[mplCg]{}) and $K=\ensuremath{\pm}2$ (\linedashdotted[mplCr]{}). The convolution (\lineS[mplCgray]{}) uses Lorentzians with a width of $\ensuremath{\Gamma}_\lo=1\:\MeV$.}
    \label{rpaImp}
\end{figure*}

\begin{figure*}[t]
    \centering
    \includegraphics{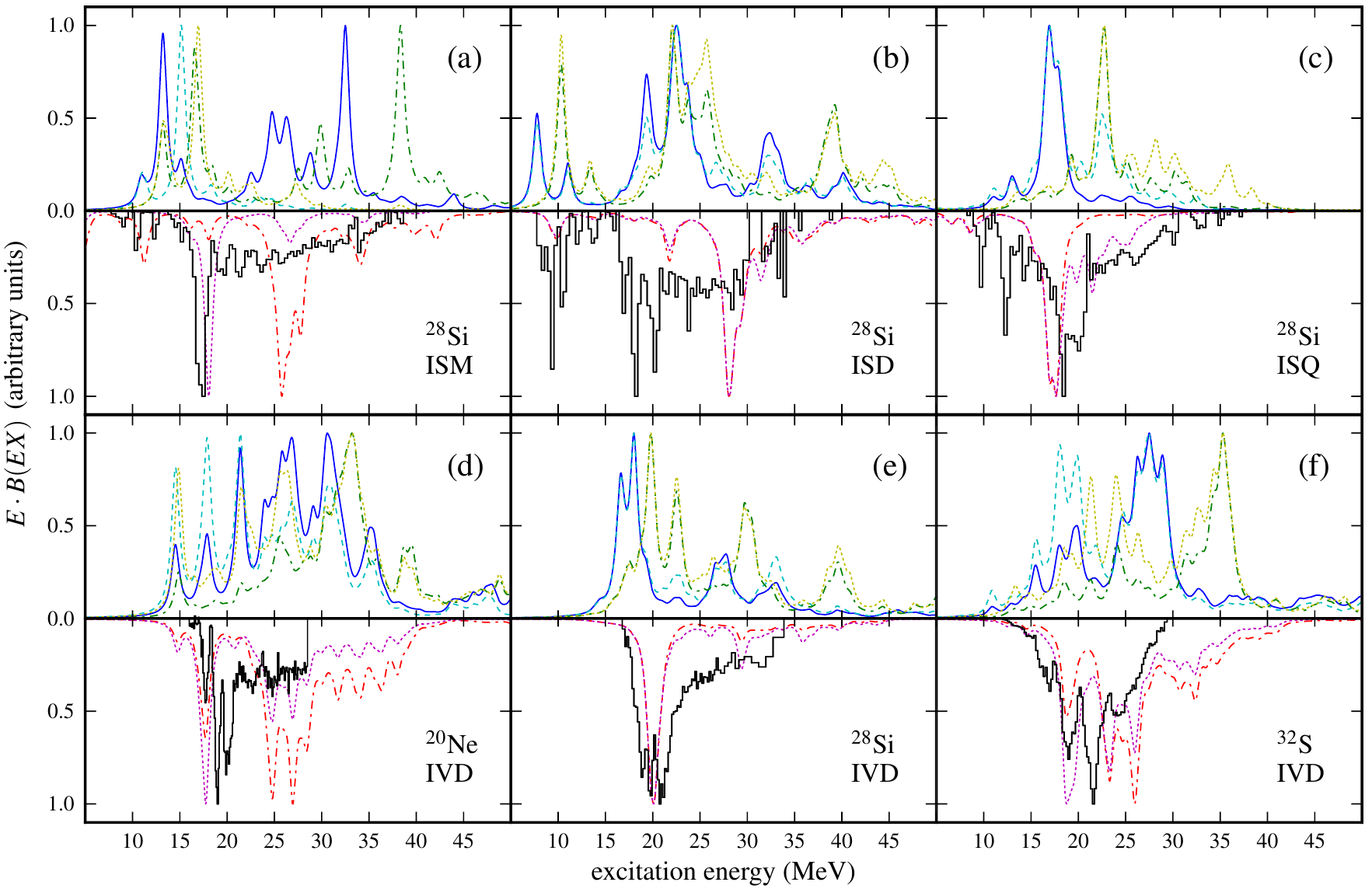}
    \caption{(color online) Comparison of selected RPA results to measured response functions. The experimental data shown as black histograms are taken from Refs.~\cite{NSR1981AL05} (\nNe IVD), \cite{NSR1983PY01} (\nSi IVD), \cite{NSR1978VA15} (\nS IVD) and \cite{Youngblood02} (\nSi isoscalar). The folded RPA response ($\ensuremath{\Gamma}_\lo=1\:\MeV$) is shown as colored curves for different interactions with and without projection: projected \SU (\lineS[mplCb]{}), intrinsic \SU (\linedashed[mplCc]{}), projected \UV (\linedashdotted[mplCg]{}), intrinsic \UV (\linedotted[mplCy]{}), projected Gogny D1S (\linedashdotted[mplCr]{}) and intrinsic Gogny D1S (\linedotted[mplCm]{}).}
    \label{rpaExp}
\end{figure*}

\begin{figure*}[t]
    \includegraphics{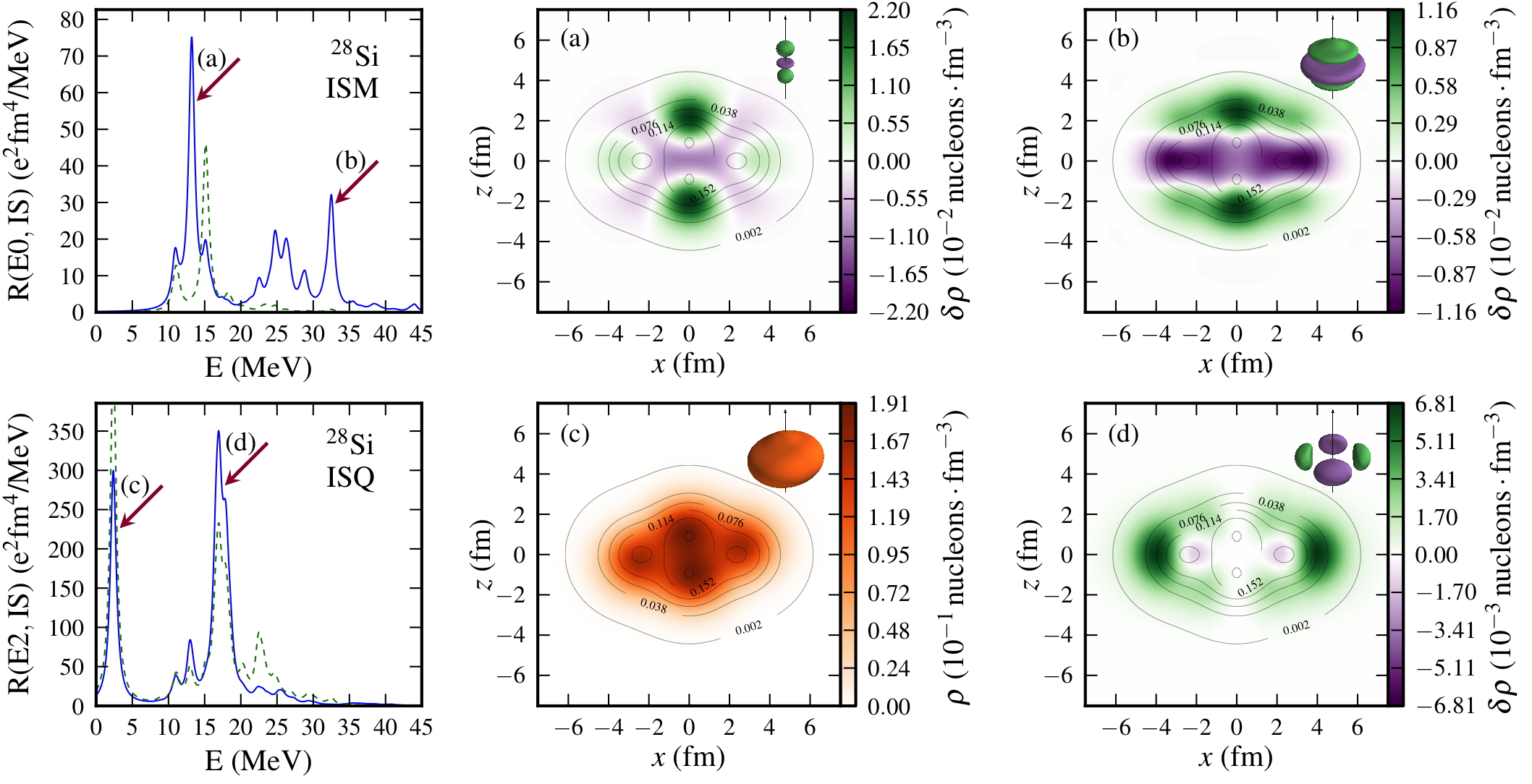}
    \caption{(color online) The left column shows the angular-momentum projected (\lineS[mplCb]{}) and intrinsic (\linedashed[mplCg]{}) ISM and ISQ response functions for \nSi obtained with the \SU interaction. The middle and right columns depict  transition densities for selected excitations, marked by the arrows in the response plots: (a) transition density for the ISM state at $\approx 14\:\MeV$, (b) transition density for the ISM state at $\approx 33\:\MeV$, (c) total density $\ensuremath{\rho}_\text{gs}+\ensuremath{\delta}\ensuremath{\rho}$ of the spurious ISQ state at $\approx 3\:\MeV$, and (d) transition density of the ISQ state at $\approx 17\:\MeV$. Contour lines (at $1\%$, $20\%$, $40\%$, $60\%$, $80\%$ and $99\%$ of $\ensuremath{\rho}_\text{max}$) always show the ground-state density for orientation. The insets show 3D isodensity surfaces of the respective (transition) density at $40\%$ of the maximum.}
    \label{rpaRho}
\end{figure*}

We start the presentation of the RPA results with a discussion of the model-space convergence and the sensitivity to the input interaction. Figure \ref{convEaHO} shows the convergence with respect to oscillator length and basis truncation for the example of the isoscalar monopole response of \nSi.
For ease of presentation, the discrete RPA transition strengths in this and the following figures are folded with  Lorentzians of width $\ensuremath{\Gamma}_\lo = 1\:\MeV$. For energies below $\approx 15\:\MeV$, all curves lie on top of each other.
In the giant resonance region at $20\text{--}30\:\MeV$, the finer details of the response differ, but the centroid of the resonance is well converged.
Above $30\:\MeV$, the differences increase, but the very prominent peak above $30\:\MeV$ is present in all calculations. We conclude that the standard basis size of $e_{\max}=14$ warrants a sufficient degree of convergence for the following discussions. The picture is similar for other response functions and nuclei.

As a second aspect we study the sensitivity of the response to the input interaction. Figure \ref{convME} again shows the isoscalar monopole response of \nSi for all interactions used in this work.
The pure two-body interaction \UV yields a rather different response than both two- plus three-body interactions, \SS and \SU.
It only shares the general features, i.e., the multi-peak structure, followed by a smaller resonance, followed by a high-energy peak, but on a stretched energy scale. This stretching can be understood in connection to the ground-state radii, which are underestimated by this interaction. In a simple mean-field picture, too small radii entail larger spacings of the single-particle levels and thus a shift of the unperturbed response to higher excitation energies \cite{Paar06}. 
The three-body part of the \SS and \SU interactions corrects for the description of the ground-state radii and leads to  smaller energy spacings of the HF energy-levels near the Fermi level, which in turn lowers the excitation energies of the collective peaks.
Although the two three-body interactions \SS and \SU yield different HF ground-state energies (cf. Sec.~\ref{hf}), the response functions are very similar and only differ in details.
This shows that the HF ground-state energy has little impact on the response functions.
In the following, we limit the presentation to the \SU interaction.
The results obtained with the Gogny D1S interaction are quite different from the other three, but since it constitutes a completely different approach, this is not unexpected.

\subsection{Structure of the collective response}

In a next step we survey the response for the standard collective modes and discuss the role of deformation in more detail. 
Figure \ref{rpaImp} shows the RPA response of the isoscalar monopole (ISM), isoscalar dipole (ISD), isovector dipole (IVD) and isoscalar quadrupole (ISQ) transition operators for \nNe, \nSi and \nS, calculated with the \SU interaction.
The discrete response in Fig. \ref{rpaImp} is color-coded to identify the different $K$-components.
Since all states with $K\neq 0$ are twofold degenerate, the corresponding lines are doubled in height.
The continuous curves again result from folding the discrete strengths with Lorentzians of width $\ensuremath{\Gamma}_\lo = 1\:\MeV$. Note that this width might be very different from the actual escape width, which cannot be described in our RPA approach. 

In axial symmetric nuclei, different oscillation modes can be classified by the projection quantum number $K$ of the total angular-momentum and the parity \ensuremath{\Pi}.
Oscillations with $(K,\ensuremath{\Pi})=(0,+)$ are along the symmetry axis and preserve the axial symmetry.
For the cases studied in this work, they appear as ISM breathing modes, ISQ \ensuremath{\beta}-vibrations or as a mixture of both.
Modes with $(0,-)$ and $(1,-)$ only appear for dipole transitions.
As the $(0,+)$ mode, the $(0,-)$ mode is along the symmetry axis and preserves axial symmetry.
However, due to the negative parity, the density increases in one half of the nucleus, while it decreases in the other.
The $(1,-)$ mode shows a similar oscillation pattern, but is directed perpendicular to the symmetry axis.
Therefore, it breaks the axial symmetry.
Spurious center-of-mass motion can appear in both types of dipole oscillations. Together with the already mentioned $(0,+)$ mode, the $(1,+)$ and the $(2,+)$ modes make up ISQ transitions.
The $(2,+)$ modes are \ensuremath{\gamma}-vibrations and the $(1,+)$ modes are rotational.
As can be seen in Fig. \ref{rpaImp}, rotational modes do not contribute much to the ISQ response, but as we will discuss later, a spurious mode can appear.

In prolate nuclei, like \nNe and \nS, the symmetry axis of the nucleus is also the longest axis.
Therefore, in the mean-field picture, $K=0$ oscillations see a shallow potential and appear at lower energy than modes with higher $K$.
This behavior is most pronounced for IVD and ISQ transitions.
IVD transitions below $20\:\MeV$ are dominated by $K=0$ modes, while those above $20\:\MeV$ are dominated by $K=1$ modes.
The ISQ giant resonance consists almost exclusively of $K=2$ modes, and, therefore, can be found at a rather high energy of $20\text{--}25\:\MeV$.

In oblate nuclei, like \nSi, the symmetry axis of the nucleus is the shortest axis.
Therefore, the situation is the exact opposite to the one found in prolate nuclei.
The low-lying IVD strength is made up exclusively by $K=1$ modes, while $K=0$ modes can only be found at very high energies.
The ISQ giant resonance is found at a significantly lower energy of $15\text{--}20\:\MeV$.

\subsection{Comparison to experiment and Gogny D1S interaction}

We can now compare the RPA response to experimental data.
In Fig. \ref{rpaExp}, the isovector dipole response is compared to data from photonuclear experiments \cite{NSR1981AL05,NSR1983PY01,NSR1978VA15} and the isoscalar response is compared to data from \ensuremath{\alpha}-scattering \cite{Youngblood02}.
Since the cross-section for dipole transitions is proportional to $E\cdot B(E1)$, the response is multiplied with the energy to ease comparison.
The measurements from \cite{Youngblood02} are given as the fraction of the energy-weighted sum-rule, so the same scaling applies also for isoscalar transitions.
As we do not consider absolute values of the transition strengths, all strength plots are normalized to the range from $0$ to $1$.

Since the \SS and \SU interactions yield almost identical results, we only show data for the \UV and \SU interactions.
For both interactions, the qualitative features of the response generally agree well with measurement, however, the exact energy of the peaks is not reproduced.

For ISM transitions, Fig. \ref{rpaExp}a, \ensuremath{\alpha}-scattering shows a strong peak between $15$ and $20\:\MeV$.
While the \UV interaction reproduces the position of this peak very well, the \SU interaction yields a peak at $\approx 14\:\MeV$.
At energies from $20\text{--}35\:\MeV$, the \UV and \SU interactions show strength, structured into multiple peaks.
In this area, experiment shows a shallow and broad structure without any distinct peaks.
As we can see in the figure, monopole strength at high energies is linked to angular-momentum projection.
Without the angular momentum-projection, there would be no strength above $25\:\MeV$.
The projected response function obtained with the Gogny D1S interaction shows almost no low-lying strength, but a very pronounced peak at $25\text{--}30\:\MeV$. In contrast, the intrinsic response reproduces the measured peak very well.

The measured ISD response, Fig. \ref{rpaExp}b, shows a few narrow peaks around $10\:\MeV$, and a broad structure from $17$ to $35\:\MeV$, with significant peaks around $20\:\MeV$.
The \UV and \SU interactions reproduce this structure, especially for the \SU interaction, the agreement with experiment is remarkably good.
Results from calculations done with the Gogny interaction do not reproduce the measured results.

For the ISQ response, Fig. \ref{rpaExp}c, the measurement shows three peaks with increasing height in the range between $10$ and $20\:\MeV$, followed by some strength up to $30\:\MeV$.
All interactions, including Gogny D1S, reproduce this shape.
Depending on the interaction, it is found lower than the measured peak (\SU and Gogny D1S) or at higher energy (\UV).

In the case of IVD transitions in \nNe, Fig. \ref{rpaExp}d, data is only available in an energy window from $16$ to $28\:\MeV$.
Measurement shows three narrow peaks at and below $20\:\MeV$, followed by a continuum up to the highest measured energy.
Our calculations for the \UV and \SU interactions show strength distributed from $15$ to above $35\:\MeV$, with more distinct peaks around $20\:\MeV$.
This is in general agreement with the measured data.
Results for the Gogny D1S interaction look a little different, but also agree with the measured data.
In the case of \nSi, Fig. \ref{rpaExp}e, we get very similar curves for both, measurement and calculation.
Here, the \UV and \SU interactions show a double-peak in the region of $20\:\MeV$, followed by strength up to $\approx30\:\MeV$.
Experiment shows a double-peak with an energy between the calculated values of the UCOM interactions.
The Gogny D1S interaction predicts a peak exactly at the measured energy, however, it is much too narrow.
For \nS, Fig. \ref{rpaExp}f, experiment shows a broad peak from $15\:\MeV$ to about $30\:\MeV$.
This is reproduced by all interactions. We do not find any sizable effects of the axial-symmetric approximation for \nS.

In conclusion, the \SU interaction, including a phenomenological three-body interaction yields a good overall agreement with the experimental response. The agreement is at the same level or sometimes better than for the purely phenomenological Gogny D1S interaction. Considering that the RPA is only a first-order approximation, the agreement with experiment is remarkable.

Motivated by this observation, we will present a detailed comparison of deformed RPA calculations for the \SU interaction with new high-resolution experiments for the IVD response in a joint publication with the experimental groups \cite{Fearick14}. There we analyze, in particular, the fine structure of the giant dipole resonance and elucidate the role of deformation driven by $\alpha$-clustering through the confrontation of high-resolution data with our calculations.

\subsection{Transition densities}

Going beyond the response, we can compute the transition densities for various discrete RPA states in order to get an intuitive geometrical understanding of the dominant excitation modes. In Fig. \ref{rpaRho}, we show the intrinsic transition densities for a few selected transitions of the ISM (a and b) and ISQ response (c and d). The figures on the left show the angular-momentum projected and the intrinsic response. For the ISM response, we see a large effect of the angular-momentum projection, while the effect on the ISQ response is rather limited.

The strong effect on the ISM response is the consequence of a mixing of breathing oscillations and \ensuremath{\beta}-vibrations through the angular-momentum projection. The projection generates a superposition of all possible rotations of the system, weighted by the angle-dependent Wigner functions to construct a predefined angular momentum.
In the case of monopole transitions, the Wigner function is a constant.
Therefore, an intrinsic \ensuremath{\beta}-vibration, like the one at $\approx33\:\MeV$, Fig. \ref{rpaRho}b, (and to some extent the state at $\approx14\:\MeV$, Fig. \ref{rpaRho}a), is converted to a monopole-type breathing oscillation by the angular-momentum projection.
This results in a redistribution of strength from the ISQ to the ISM channel for these states, which can best be seen for the state at $\approx33\:\MeV$.
In this case, the intrinsic ISM transition strength practically vanishes, whereas the angular-momentum projected strength provides the second strongest peak in the whole response.

At this point, a comment is in order on the so-called needle approximation for the angular-momentum projection, which is used, e.g., in Refs.~\cite{Arteaga08,Peru08}.
For monopole transitions, the needle approximation simply reproduces the intrinsic response and, thus, misses some major effects of the projection, as can be seen in Fig. \ref{rpaRho}.

Another interesting state is the ISQ state at $\approx3\:\MeV$, Fig. \ref{rpaRho}c, which has $K=1$.
This state appears for all deformed nuclei in the ISQ response, but is not shown in Fig. \ref{rpaImp} and \ref{rpaExp}.
The angular-momentum projection strongly reduces the strength of this state.
Further investigation shows, that the state has very large $Y$-amplitudes---about the same order of magnitude as the $X$-amplitudes.
This suggest a spurious rotational state, as is expected for deformed nuclei, which is confirmed by the transition density.
For this state, Fig. \ref{rpaRho}c shows not the transition density, but the total density $\ensuremath{\rho}_\text{gs}+\ensuremath{\delta}\ensuremath{\rho}$.
The shape of the nucleus remains unchanged and it is only rotated around the $y$-axis.
As this spurious mode is found at an energy significantly above zero, it can contaminate other, non-spurious states with $(K,\ensuremath{\Pi})=(1,+)$.
However, since these modes do not contribute significantly to the ISQ response, this does not pose a problem for the current studies.
Spurious center-of-mass modes are also found for the $(0,-)$ and $(1,-)$ modes, but lie at zero energy.

As was already seen in Fig. \ref{rpaImp}, the ISQ response is dominated by $K=2$ transitions, which correspond to \ensuremath{\gamma}-vibrations.
Figure \ref{rpaRho}d shows the transition density of the strongest ISQ state, which is indeed a perfect \ensuremath{\gamma}-vibration.

\section{Conclusions}

In this paper, HF and RPA calculations with unitarily transformed realistic interactions for axially-symmetric deformed nuclei have been carried out for the first time.
To obtain ground-state energies and response functions in the lab-frame, an explicit angular-momentum projection was employed.

For all studied nuclei, we find a much stronger fragmentation of the resonances than in spherical nuclei.
Due to the angular-momentum projection, the ISM response extends to energies as high as $40\:\MeV$.
For the IVD response, we find the expected dipole splitting.
In prolate nuclei, oscillations along the symmetry axis are found at lower energies, while those perpendicular to the symmetry axis are found at higher energies.
In oblate nuclei, the situation is reversed.
In case of the ISD response, we find multiple peaks at high and low energies.
The ISQ response is clearly dominated by \ensuremath{\gamma}-vibrations.
The geometry of the individual oscillation modes was studied via transition densities, which confirmed spurious rotational states and the effect of the angular-momentum projection on breathing oscillations and \ensuremath{\beta}-vibrations.

In comparison to experiment, the unitarily transformed interactions, in particular the \SU and \SS interactions, which also yield the correct radius systematics, provide a good overall description of the major collective modes in our deformed and angular-momentum projected RPA calculations. The quality of the agreement is comparable to, and sometimes better than results obtained with phenomenological interactions, such as the Gogny D1S interaction. This already shows that a good reproduction of the ground-state energies at the HF level is neither a necessary nor a sufficient criterion for a good description of the collective response. 

This study opens multiple lines of research for the future. Motivated by the good agreement with experiment and the fact that significant fragmentation is already present in the RPA response, we will analyze the fine structure of the giant dipole resonance and compare to new high-resolution data for the nuclei discussed \cite{Fearick14}. This will shed light on the role of deformation and $\alpha$-clustering on the fragmentation and fine structure of giant resonances.
Another obvious extension of the present work is the use of two- plus three-nucleon interactions from chiral EFT. First studies along these lines with a spherical formulation of the RPA are well advanced. However, present chiral interactions, even after the inclusion of the chiral three-nucleon terms, still underestimate the radii of medium-mass nuclei \cite{Binder14}. It remains to be seen how consistent next-generation chiral Hamiltonians will behave in this respect.
Finally, the extension from first- to second-order RPA will be a target of future studies. As a first step, second-order RPA calculations for spherical nuclei including realistic 3N interactions are already under way \cite{Trippel14}.

\section*{Acknowledgements}
This work was supported by the Deutsche Forschungsgemeinschaft through SFB 634, by the Helmholtz International Center for FAIR within the framework of the LOEWE program launched by the state of Hesse, and the German Federal Ministry of Education and Research (BMBF 06DA9040I, 06DA7047I).

\appendix
\section{Angular-momentum projected transition amplitudes in the RPA framework}  \label{appSTR}
The unprojected transition amplitudes to the RPA ground state are obtained by applying the quasi-boson approximation \cite{Rowe70}, followed by straightforward calculation
\begin{equation}
\begin{split}
    \ensuremath{\langle}\rpa|&\op{T}_{\ensuremath{\lambda}\ensuremath{\mu}}\,\Qc|\rpa\ensuremath{\rangle}\approx\ensuremath{\langle}\hf|[\op{T}_{\ensuremath{\lambda}\ensuremath{\mu}},\Qc]|\hf\ensuremath{\rangle} \\ 
    &= \sum_{ma} \left( X^\ensuremath{\omega}_{ma}\, \ensuremath{\langle}a|\op{T}_{\ensuremath{\lambda}\ensuremath{\mu}}|m\ensuremath{\rangle} + Y^\ensuremath{\omega}_{ma}\, \ensuremath{\langle}m|\op{T}_{\ensuremath{\lambda}\ensuremath{\mu}}|a\ensuremath{\rangle} \right) \;.
\end{split}
\end{equation}
It would be desirable to derive the projected RPA transition amplitudes in a similar manner, directly from the equation for projected transition amplitudes \cite{Ring80}
\begin{equation} \mlabel{rpa:str-proj}
\begin{split}
    (J_0\|\op{T}^\ensuremath{\lambda}\|J_\ensuremath{\omega})&=(2\,J_0+1) \, N_0 \, N_\ensuremath{\omega} \,  \sum_{\substack{K_0 K_\ensuremath{\omega}\\\ensuremath{\mu}}} \,g^{(0)}_{K_0}\,g^{(\ensuremath{\omega})}_{K_\ensuremath{\omega}}\\ 
    &\times (-1)^{\ensuremath{\lambda}+J_\ensuremath{\omega}+K_0}  \mat{J_0 & J_\ensuremath{\omega} & \ensuremath{\lambda} \\ -K_0 & K_0-\ensuremath{\mu} & \ensuremath{\mu}}\\ 
    &\times \ensuremath{\langle}\rpa|\op{T}_{\ensuremath{\lambda}\ensuremath{\mu}}\, \op{P}^{J_\ensuremath{\omega}}_{K_0-\ensuremath{\mu}, K_\ensuremath{\omega}}\,\op{Q}^\ensuremath{\dagger}_\ensuremath{\omega}|\rpa\ensuremath{\rangle} \;.
\end{split}
\end{equation}
However, this is not possible in a consistent and unambiguous way.
The canonical way of the RPA is to replace pairs of operators with their commutators and the RPA ground-state with the HF ground-state.
This treatment is not possible because of the projection operator.
Since the projection operator projects a fixed set of quantum numbers onto another fixed set, these quantum numbers would have to change according to the order of the operators $\op T^\ensuremath{\lambda}_\ensuremath{\mu}$ and $\Qc$---otherwise the projection operator would annihilate the states.
One could relax the requirement of a commutator and allow anything to be added that vanishes for the true RPA states, but still gives a $Y$-amplitude contribution in the QBA.
Then, the quantum numbers of the projection operator could be changed to match the other operators.
However, this scheme also allows the introduction of an arbitrary phase---and this freedom has to be exploited if one wants to reproduce the unprojected results for spherical nuclei.
But since this treatment is ambiguous and leaves much to be desired in terms of simplicity, we opt for a different, less ambiguous approach.

To calculate the projected transition amplitudes, we again consider the unprojected intrinsic transition amplitudes.
The complete transition amplitude of multipolarity \ensuremath{\lambda} including normalization factors is given by
\begin{equation}
    \begin{split}
        \frac{\ensuremath{\langle}\rpa|\op{T}^\ensuremath{\lambda}|\ensuremath{\omega}\ensuremath{\rangle}}{\sqrt{\ensuremath{\langle}\rpa|\rpa\ensuremath{\rangle}\ensuremath{\langle}\ensuremath{\omega}|\ensuremath{\omega}\ensuremath{\rangle}}} &= \sum_\ensuremath{\mu} \frac{\ensuremath{\langle}\rpa|\op{T}_{\ensuremath{\lambda}\ensuremath{\mu}}|\ensuremath{\omega}\ensuremath{\rangle}}{\sqrt{\ensuremath{\langle}\rpa|\rpa\ensuremath{\rangle}\ensuremath{\langle}\ensuremath{\omega}|\ensuremath{\omega}\ensuremath{\rangle}}} \\
        &= \sum_\ensuremath{\mu} \frac{\ensuremath{\langle}\rpa|\op{T}_{\ensuremath{\lambda}\ensuremath{\mu}}\,\Qc|\rpa\ensuremath{\rangle}}{\sqrt{\ensuremath{\langle}\rpa|\rpa\ensuremath{\rangle}\ensuremath{\langle}\rpa|\Qd\,\Qc|\rpa\ensuremath{\rangle}}} \;.
    \end{split}
\end{equation}
The numerator is given by Eq. \eqref{rpa:rpa-int}
\begin{equation}
    \sum_\ensuremath{\mu} \ensuremath{\langle}\hf|[\op{T}_{\ensuremath{\lambda}\ensuremath{\mu}},\Qc]|\hf\ensuremath{\rangle} 
    = \sum_{\ensuremath{\mu},ma} \left( X^\ensuremath{\omega}_{ma}\, \ensuremath{\langle}a|\op{T}_{\ensuremath{\lambda}\ensuremath{\mu}}|m\ensuremath{\rangle} + Y^\ensuremath{\omega}_{ma}\, \ensuremath{\langle}m|\op{T}_{\ensuremath{\lambda}\ensuremath{\mu}}|a\ensuremath{\rangle} \right) \;.
\end{equation}
Assuming real matrix elements, we write the solution in a form that is more suitable for use in the projected formalism

\vspace*{2ex}
\;

\begin{equation} \mlabel{rpa:str-int}
\begin{split}
    &\sum_\ensuremath{\mu} \ensuremath{\langle}\hf|[\op{T}_{\ensuremath{\lambda}\ensuremath{\mu}},\Qc]|\hf\ensuremath{\rangle}\\ 
    &= \sum_{\ensuremath{\mu},ma} \big( X^\ensuremath{\omega}_{ma}\, \ensuremath{\langle}\hf|\op{T}_{\ensuremath{\lambda}\ensuremath{\mu}}\,\ac_m\ad_a|\hf\ensuremath{\rangle}
    + Y^\ensuremath{\omega}_{ma}\,(-1)^\ensuremath{\mu} \ensuremath{\langle}\hf|\op{T}_{\ensuremath{\lambda}-\ensuremath{\mu}}\,\ac_m\ad_a|\hf\ensuremath{\rangle} \big) \\
    &=\sum_{ma} \left( X^\ensuremath{\omega}_{ma} + (-1)^{K_{ma}}Y^\ensuremath{\omega}_{ma}\right)\,\sum_\ensuremath{\mu} \, \ensuremath{\langle}\hf|\, \op{T}_{\ensuremath{\lambda}\ensuremath{\mu}}\,\ac_m\ad_a|\hf\ensuremath{\rangle} \;.
\end{split}
\end{equation}
We renamed $-\ensuremath{\mu}$ to $\ensuremath{\mu}$ in the $Y$-amplitude part and used that $K$ is a well defined quantum number in axially symmetric nuclei.
$K_{ma}$ denotes the $K$ quantum number of the state $\ac_m\ad_a|\hf\ensuremath{\rangle}$.

To get the projected transition amplitudes, we simply include the $XY$-factor in the formula for the projected transition amplitudes \eqref{rpa:str-proj}
\begin{equation} \mlabel{rpa:rpa-proj}
\begin{split}
    (\rpa\|&\op{T}^\ensuremath{\lambda}\|\ensuremath{\omega}) = (2\,J_0+1) \, N_0 \, N_\ensuremath{\omega} \,  (-1)^{J_0-K_0} \\
    &\times \sum_{ma} \left( X^\ensuremath{\omega}_{ma} + (-1)^{K_{ma}}Y^\ensuremath{\omega}_{ma}\right) \, \sum_{\ensuremath{\mu}} \mat{J_0 & \ensuremath{\lambda} & J_\ensuremath{\omega} \\ -K_\hf & \ensuremath{\mu} & K_\hf-\ensuremath{\mu}} \\
    &\times\ensuremath{\langle}\hf|\op{T}_{\ensuremath{\lambda}\ensuremath{\mu}}\, \op{P}^{J_\ensuremath{\omega}}_{K_\hf-\ensuremath{\mu}, K_\hf+K_{ma}}\ac_m\ad_a|\hf\ensuremath{\rangle} \;.
\end{split}
\end{equation}
Simplifications arising from axial symmetry have already been applied.
We treat the normalization factors accordingly.
The normalization factor from the RPA ground state is given by
\begin{equation} \mlabel{rpa:normRPAgs-as}
    N_0 = \sqrt{\ensuremath{\langle}\hf|\op{P}^{J_0}_{K_\hf,K_\hf}|\hf\ensuremath{\rangle}}^{-1} \;.
\end{equation}
The unprojected normalization factor for the excited state $N_\ensuremath{\omega}$ is given by
\begin{equation}
\begin{split}
    \ensuremath{\langle}\rpa|\Qd\,\Qc|\rpa\ensuremath{\rangle} &\approx \ensuremath{\langle}\hf|[\Qd ,\Qc]|\hf\ensuremath{\rangle} \\
    &= \sum_{ma} \left( X^\ensuremath{\omega}_{ma}\,X^\ensuremath{\omega}_{ma} - Y^\ensuremath{\omega}_{ma}\,Y^\ensuremath{\omega}_{ma} \right) \;, 
\end{split}
\end{equation}
which evaluates to unity due to the orthonormality of the RPA amplitudes.
We, therefore, use the following projected normalization factor $N_\ensuremath{\omega}$
\begin{equation} \mlabel{rpa:normRPAex-as}
\begin{split}
    N_\ensuremath{\omega}^{-2} &= \sum_{ma,nb} \left( X^\ensuremath{\omega}_{ma} \,X^\ensuremath{\omega}_{nb} - Y^\ensuremath{\omega}_{ma}\,Y^\ensuremath{\omega}_{nb}\right)\\
    &\times \ensuremath{\langle}\hf|\ac_m\ad_a \, \op{P}^{J_\ensuremath{\omega}}_{K_\hf+k_{am}, K_\hf+k_{nb}}\ac_n\ad_b|\hf\ensuremath{\rangle} \;.
\end{split}
\end{equation}

%


\begin{thebibliography}{10}%
\makeatletter
\providecommand \@ifxundefined [1]{%
 \ifx #1\undefined \expandafter \@firstoftwo
 \else \expandafter \@secondoftwo
\fi
}%
\providecommand \@ifnum [1]{%
 \ifnum #1\expandafter \@firstoftwo
 \else \expandafter \@secondoftwo
\fi
}%
\providecommand \enquote [1]{``#1''}%
\providecommand \bibnamefont  [1]{#1}%
\providecommand \bibfnamefont [1]{#1}%
\providecommand \citenamefont [1]{#1}%
\providecommand\href[0]{\@sanitize\@href}%
\providecommand\@href[1]{\endgroup\@@startlink{#1}\endgroup\@@href}%
\providecommand\@@href[1]{#1\@@endlink}%
\providecommand \@sanitize [0]{\begingroup\catcode`\&12\catcode`\#12\relax}%
\@ifxundefined \pdfoutput {\@firstoftwo}{%
 \@ifnum{\z@=\pdfoutput}{\@firstoftwo}{\@secondoftwo}%
}{%
 \providecommand\@@startlink[1]{\leavevmode}%
 \providecommand\@@endlink[0]{}%
}{%
 \providecommand\@@startlink[1]{%
  \leavevmode
  \pdfstartlink
   attr{/Border[0 0 1 ]/H/I/C[0 1 1]}%
   user{/Subtype/Link/A<</Type/Action/S/URI/URI(#1)>>}%
  \relax
 }%
 \providecommand\@@endlink[0]{\pdfendlink}%
}%
\providecommand \url  [0]{\begingroup\@sanitize \@url }%
\providecommand \@url [1]{\endgroup\@href {#1}{\urlprefix}}%
\providecommand \urlprefix [0]{URL }%
\providecommand \Eprint[0]{\href }%
\@ifxundefined \urlstyle {%
  \providecommand \doi [1]{doi:\discretionary{}{}{}#1}%
}{%
  \providecommand \doi [0]{doi:\discretionary{}{}{}\begingroup
  \urlstyle{rm}\Url }%
}%
\providecommand \doibase [0]{http://dx.doi.org/}%
\providecommand \Doi[1]{\href{\doibase#1}}%
\providecommand \bibAnnote [3]{%
  \BibitemShut{#1}%
  \begin{quotation}\noindent
    \textsc{Key:}\ #2\\\textsc{Annotation:}\ #3%
  \end{quotation}%
}%
\providecommand \bibAnnoteFile [2]{%
  \IfFileExists{#2}{\bibAnnote {#1} {#2} {\input{#2}}}{}%
}%
\providecommand \typeout [0]{\immediate \write \m@ne }%
\providecommand \selectlanguage [0]{\@gobble}%
\providecommand \bibinfo [0]{\@secondoftwo}%
\providecommand \bibfield [0]{\@secondoftwo}%
\providecommand \translation [1]{[#1]}%
\providecommand \BibitemOpen[0]{}%
\providecommand \bibitemStop [0]{}%
\providecommand \bibitemNoStop [0]{.\EOS\space}%
\providecommand \EOS [0]{\spacefactor3000\relax}%
\providecommand \BibitemShut [1]{\csname bibitem#1\endcsname}%
\bibitem{Rowe70}%
  \BibitemOpen
  \bibfield{author}{%
  \bibinfo {author} {\bibfnamefont{D.}~\bibnamefont{Rowe}},\ }%
  \emph{\bibinfo {title} {Nuclear Collective Motion}}\ (\bibinfo {publisher}
  {Methuen and Co. Ltd.},\ \bibinfo {year} {1970})%
  \bibAnnoteFile{NoStop}{Rowe70}%
\bibitem{Skyrme58}%
  \BibitemOpen
  \bibfield{author}{%
  \bibinfo {author} {\bibfnamefont{T.}~\bibnamefont{Skyrme}},\ }%
  \bibfield{journal}{%
  \Doi{10.1016/0029-5582(58)90345-6}{\bibinfo {journal} {Nucl. Phys.}}\ }%
  \textbf{\bibinfo {volume} {9}},\ \bibinfo {pages} {615} (\bibinfo {year}
  {1958})
  \bibAnnoteFile{NoStop}{Skyrme58}%
\bibitem{Chabanat98}%
  \BibitemOpen
  \bibfield{author}{%
  \bibinfo {author} {\bibfnamefont{E.}~\bibnamefont{Chabanat}}, \bibinfo
  {author} {\bibfnamefont{P.}~\bibnamefont{Bonche}}, \bibinfo {author}
  {\bibfnamefont{P.}~\bibnamefont{Haensel}}, \bibinfo {author}
  {\bibfnamefont{J.}~\bibnamefont{Meyer}},\ and\ \bibinfo {author}
  {\bibfnamefont{R.}~\bibnamefont{Schaeffer}},\ }%
  \bibfield{journal}{%
  \Doi{10.1016/S0375-9474(98)00180-8}{\bibinfo {journal} {Nucl. Phys. A}}\
  }%
  \textbf{\bibinfo {volume} {635}},\ \bibinfo {pages} {231} (\bibinfo {year}
  {1998})
  \bibAnnoteFile{NoStop}{Chabanat98}%
\bibitem{Tondeur00}%
  \BibitemOpen
  \bibfield{author}{%
  \bibinfo {author} {\bibfnamefont{F.}~\bibnamefont{Tondeur}}, \bibinfo
  {author} {\bibfnamefont{S.}~\bibnamefont{Goriely}}, \bibinfo {author}
  {\bibfnamefont{J.~M.}\ \bibnamefont{Pearson}},\ and\ \bibinfo {author}
  {\bibfnamefont{M.}~\bibnamefont{Onsi}},\ }%
  \bibfield{journal}{%
  \Doi{10.1103/PhysRevC.62.024308}{\bibinfo {journal} {Phys. Rev. C}}\ }%
  \textbf{\bibinfo {volume} {62}},\ \bibinfo {pages} {024308} (\bibinfo {year} {2000})\
  \bibAnnoteFile{NoStop}{Tondeur00}%
\bibitem{DG80}%
  \BibitemOpen
  \bibfield{author}{%
  \bibinfo {author} {\bibfnamefont{J.}~\bibnamefont{Decharg{\'e}}}\ and\
  \bibinfo {author} {\bibfnamefont{D.}~\bibnamefont{Gogny}},\ }%
  \bibfield{journal}{%
  \Doi{10.1103/PhysRevC.21.1568}{\bibinfo {journal} {Phys. Rev. C}}\ }%
  \textbf{\bibinfo {volume} {21}},\ \bibinfo {pages} {1568} (\bibinfo {year} {1980})
  \bibAnnoteFile{NoStop}{DG80}%
\bibitem{Wiringa95}%
  \BibitemOpen
  \bibfield{author}{%
  \bibinfo {author} {\bibfnamefont{R.~B.}\ \bibnamefont{Wiringa}}, \bibinfo
  {author} {\bibfnamefont{V.~G.~J.}\ \bibnamefont{Stoks}},\ and\ \bibinfo
  {author} {\bibfnamefont{R.}~\bibnamefont{Schiavilla}},\ }%
  \bibfield{journal}{%
  \Doi{10.1103/PhysRevC.51.38}{\bibinfo {journal} {Phys. Rev. C}}\ }%
  \textbf{\bibinfo {volume} {51}},\ \bibinfo {pages} {38} (\bibinfo {year} {1995})
  \bibAnnoteFile{NoStop}{Wiringa95}%
\bibitem{Machleidt01}%
  \BibitemOpen
  \bibfield{author}{%
  \bibinfo {author} {\bibfnamefont{R.}~\bibnamefont{Machleidt}},\ }%
  \bibfield{journal}{%
  \Doi{10.1103/PhysRevC.63.024001}{\bibinfo {journal} {Phys. Rev. C}}\ }%
  \textbf{\bibinfo {volume} {63}},\ \bibinfo {pages} {024001} (\bibinfo {year} {2001})
  \bibAnnoteFile{NoStop}{Machleidt01}%
\bibitem{Stoks94}%
  \BibitemOpen
  \bibfield{author}{%
  \bibinfo {author} {\bibfnamefont{V.~G.~J.}\ \bibnamefont{Stoks}}, \bibinfo
  {author} {\bibfnamefont{R.~A.~M.}\ \bibnamefont{Klomp}}, \bibinfo {author}
  {\bibfnamefont{C.~P.~F.}\ \bibnamefont{Terheggen}},\ and\ \bibinfo {author}
  {\bibfnamefont{J.~J.}\ \bibnamefont{de~Swart}},\ }%
  \bibfield{journal}{%
  \Doi{10.1103/PhysRevC.49.2950}{\bibinfo {journal} {Phys. Rev. C}}\ }%
  \textbf{\bibinfo {volume} {49}},\ \bibinfo {pages} {2950} (\bibinfo {year} {1994})
  \bibAnnoteFile{NoStop}{Stoks94}%
\bibitem{Epelbaum02}%
  \BibitemOpen
  \bibfield{author}{%
  \bibinfo {author} {\bibfnamefont{E.}~\bibnamefont{Epelbaum}}, \bibinfo
  {author} {\bibfnamefont{A.}~\bibnamefont{Nogga}}, \bibinfo {author}
  {\bibfnamefont{W.}~\bibnamefont{Gl{\"o}ckle}}, \bibinfo {author}
  {\bibfnamefont{H.}~\bibnamefont{Kamada}}, \bibinfo {author}
  {\bibfnamefont{U.-G.}\ \bibnamefont{Mei{\ss}ner}},\ and\ \bibinfo {author}
  {\bibfnamefont{H.}~\bibnamefont{Wita{\l}a}},\ }%
  \bibfield{journal}{%
  \Doi{10.1103/PhysRevC.66.064001}{\bibinfo {journal} {Phys. Rev. C}}\ }%
  \textbf{\bibinfo {volume} {66}},\ \bibinfo {pages} {064001} (\bibinfo {year} {2002})
  \bibAnnoteFile{NoStop}{Epelbaum02}%
\bibitem{Entem03}%
  \BibitemOpen
  \bibfield{author}{%
  \bibinfo {author} {\bibfnamefont{D.~R.}\ \bibnamefont{Entem}}\ and\ \bibinfo
  {author} {\bibfnamefont{R.}~\bibnamefont{Machleidt}},\ }%
  \bibfield{journal}{%
  \Doi{10.1103/PhysRevC.68.041001}{\bibinfo {journal} {Phys. Rev. C}}\ }%
  \textbf{\bibinfo {volume} {68}},\ \bibinfo {pages} {041001} (\bibinfo {year} {2003})
  \bibAnnoteFile{NoStop}{Entem03}%
\bibitem{Epelbaum05}%
  \BibitemOpen
  \bibfield{author}{%
  \bibinfo {author} {\bibfnamefont{E.}~\bibnamefont{Epelbaum}}, \bibinfo
  {author} {\bibfnamefont{W.}~\bibnamefont{Gl{\"o}ckle}},\ and\ \bibinfo
  {author} {\bibfnamefont{U.-G.}\ \bibnamefont{Mei{\ss}ner}},\ }%
  \bibfield{journal}{%
  \Doi{10.1016/j.nuclphysa.2004.09.107}{\bibinfo {journal} {Nucl. Phys.
  A}}\ }%
  \textbf{\bibinfo {volume} {747}},\ \bibinfo {pages} {362} (\bibinfo {year}
  {2005})
  \bibAnnoteFile{NoStop}{Epelbaum05}%
\bibitem{Machleidt10}%
  \BibitemOpen
  \bibfield{author}{%
  \bibinfo {author} {\bibfnamefont{R.}~\bibnamefont{Machleidt}}\ and\ \bibinfo
  {author} {\bibfnamefont{D.~R.}\ \bibnamefont{Entem}},\ }%
  \bibfield{journal}{%
  \bibinfo {journal} {J. Phys. G: Nucl. Part. Phys.}\ }%
  \textbf{\bibinfo {volume} {37}},\ \bibinfo {pages} {064041} (\bibinfo {year}
  {2010})
  \bibAnnoteFile{NoStop}{Machleidt10}%
\bibitem{Roth10}%
  \BibitemOpen
  \bibfield{author}{%
  \bibinfo {author} {\bibfnamefont{R.}~\bibnamefont{Roth}}, \bibinfo {author}
  {\bibfnamefont{T.}~\bibnamefont{Neff}},\ and\ \bibinfo {author}
  {\bibfnamefont{H.}~\bibnamefont{Feldmeier}},\ }%
  \bibfield{journal}{%
  \Doi{DOI: 10.1016/j.ppnp.2010.02.003}{\bibinfo {journal} {Prog. 
  Part. Nucl. Phys.}}\ }%
  \textbf{\bibinfo {volume} {65}},\ \bibinfo {pages} {50} (\bibinfo {year}
  {2010})
  \bibAnnoteFile{NoStop}{Roth10}%
\bibitem{Peru08}%
  \BibitemOpen
  \bibfield{author}{%
  \bibinfo {author} {\bibfnamefont{S.}~\bibnamefont{Peru}}\ and\ \bibinfo
  {author} {\bibfnamefont{H.}~\bibnamefont{Goutte}},\ }%
  \bibfield{journal}{%
  \bibinfo {journal} {Phys. Rev. C}\ }%
  \textbf{\bibinfo {volume} {77}},\ \bibinfo {pages} {044313} (\bibinfo {year}
  {2008})%
  \bibAnnoteFile{NoStop}{Peru08}%
\bibitem{Arteaga08}%
  \BibitemOpen
  \bibfield{author}{%
  \bibinfo {author} {\bibfnamefont{D.~P.}\ \bibnamefont{Arteaga}}\ and\
  \bibinfo {author} {\bibfnamefont{P.}~\bibnamefont{Ring}},\ }%
  \bibfield{journal}{%
  \bibinfo {journal} {Phys. Rev. C}\ }%
  \textbf{\bibinfo {volume} {77}},\ \bibinfo {pages} {034317} (\bibinfo {year}
  {2008})%
  \bibAnnoteFile{NoStop}{Arteaga08}%
\bibitem{Losa10}%
  \BibitemOpen
  \bibfield{author}{%
  \bibinfo {author} {\bibfnamefont{C.}~\bibnamefont{Losa}}, \bibinfo {author}
  {\bibfnamefont{A.}~\bibnamefont{Pastore}}, \bibinfo {author}
  {\bibfnamefont{T.}~\bibnamefont{Dossing}}, \bibinfo {author}
  {\bibfnamefont{E.}~\bibnamefont{Vigezzi}},\ and\ \bibinfo {author}
  {\bibfnamefont{R.}~\bibnamefont{Broglia}},\ }%
  \bibfield{journal}{%
  \bibinfo {journal} {Phys. Rev. C}\ }%
  \textbf{\bibinfo {volume} {81}},\ \bibinfo {pages} {064307} (\bibinfo {year}
  {2010})%
  \bibAnnoteFile{NoStop}{Losa10}%
\bibitem{Loewdin55-1}%
  \BibitemOpen
  \bibfield{author}{%
  \bibinfo {author} {\bibfnamefont{P.-O.}\ \bibnamefont{L{\"o}wdin}},\ }%
  \bibfield{journal}{%
  \bibinfo {journal} {Phys. Rev.}\ }%
  \textbf{\bibinfo {volume} {97}},\ \bibinfo {pages} {1475} (\bibinfo {year}
  {1955})%
  \bibAnnoteFile{NoStop}{Loewdin55-1}%
\bibitem{Brink66}%
  \BibitemOpen
  \bibfield{author}{%
  \bibinfo {author} {\bibfnamefont{D.}~\bibnamefont{Brink}},\ }%
  in\ \emph{\bibinfo {booktitle} {Many-Body Description of Nuclear Structure
  and Reactions}},\ \bibinfo {series} {International School of Physics "Erico
  Fermi"}, Vol.~\bibinfo {volume} {36}\ (\bibinfo {year} {1966})\ pp.\ \bibinfo
  {pages} {247--277}%
  \bibAnnoteFile{NoStop}{Brink66}%
\bibitem{Ring80}%
  \BibitemOpen
  \bibfield{author}{%
  \bibinfo {author} {\bibfnamefont{P.}~\bibnamefont{Ring}}\ and\ \bibinfo
  {author} {\bibfnamefont{P.}~\bibnamefont{Schuck}},\ }%
  \emph{\bibinfo {title} {The Nuclear Many-Body Problem}},\ Texts and
  Monographs in Physics\ (\bibinfo {publisher} {Springer Berlin Heidelberg},\
  \bibinfo {year} {1980})\ ISBN \bibinfo {isbn} {978-3-540-21206-5 (Print)}%
  \bibAnnoteFile{NoStop}{Ring80}%
\bibitem{Roth05}%
  \BibitemOpen
  \bibfield{author}{%
  \bibinfo {author} {\bibfnamefont{R.}~\bibnamefont{Roth}}, \bibinfo {author}
  {\bibfnamefont{H.}~\bibnamefont{Hergert}}, \bibinfo {author}
  {\bibfnamefont{P.}~\bibnamefont{Papakonstantinou}}, \bibinfo {author}
  {\bibfnamefont{T.}~\bibnamefont{Neff}},\ and\ \bibinfo {author}
  {\bibfnamefont{H.}~\bibnamefont{Feldmeier}},\ }%
  \bibfield{journal}{%
  \Doi{10.1103/PhysRevC.72.034002}{\bibinfo {journal} {Phys. Rev. C}}\ }%
  \textbf{\bibinfo {volume} {72}},\ \bibinfo {pages} {034002} (\bibinfo {year} {2005})%
  \bibAnnoteFile{NoStop}{Roth05}%
\bibitem{Roth06}%
  \BibitemOpen
  \bibfield{author}{%
  \bibinfo {author} {\bibfnamefont{R.}~\bibnamefont{Roth}}, \bibinfo {author}
  {\bibfnamefont{P.}~\bibnamefont{Papakonstantinou}}, \bibinfo {author}
  {\bibfnamefont{N.}~\bibnamefont{Paar}}, \bibinfo {author}
  {\bibfnamefont{H.}~\bibnamefont{Hergert}}, \bibinfo {author}
  {\bibfnamefont{T.}~\bibnamefont{Neff}},\ and\ \bibinfo {author}
  {\bibfnamefont{H.}~\bibnamefont{Feldmeier}},\ }%
  \bibfield{journal}{%
  \Doi{10.1103/PhysRevC.73.044312}{\bibinfo {journal} {Phys. Rev. C}}\ }%
  \textbf{\bibinfo {volume} {73}},\ \bibinfo {pages} {044312} (\bibinfo {year} {2006})
  \bibAnnoteFile{NoStop}{Roth06}%
\bibitem{Roth08}%
  \BibitemOpen
  \bibfield{author}{%
  \bibinfo {author} {\bibfnamefont{R.}~\bibnamefont{Roth}}, \bibinfo {author}
  {\bibfnamefont{S.}~\bibnamefont{Reinhardt}},\ and\ \bibinfo {author}
  {\bibfnamefont{H.}~\bibnamefont{Hergert}},\ }%
  \bibfield{journal}{%
  {\bibinfo {journal} {Phys. Rev. C}}\ }%
  \textbf{\bibinfo {volume} {77}},\ \bibinfo {pages} {064003} (\bibinfo {year} {2008})
  \bibAnnoteFile{NoStop}{Roth08}%
\bibitem{Paar06}%
  \BibitemOpen
  \bibfield{author}{%
  \bibinfo {author} {\bibfnamefont{N.}~\bibnamefont{Paar}}, \bibinfo {author}
  {\bibfnamefont{P.}~\bibnamefont{Papakonstantinou}}, \bibinfo {author}
  {\bibfnamefont{H.}~\bibnamefont{Hergert}},\ and\ \bibinfo {author}
  {\bibfnamefont{R.}~\bibnamefont{Roth}},\ }%
  \bibfield{journal}{%
  \Doi{10.1103/PhysRevC.74.014318}{\bibinfo {journal} {Phys. Rev. C}}\ }%
  \textbf{\bibinfo {volume} {74}},\ \bibinfo {pages} {014318} (\bibinfo {year} {2006})
  \bibAnnoteFile{NoStop}{Paar06}%
\bibitem{Papakonstantinou07}%
  \BibitemOpen
  \bibfield{author}{%
  \bibinfo {author} {\bibfnamefont{P.}~\bibnamefont{Papakonstantinou}},
  \bibinfo {author} {\bibfnamefont{R.}~\bibnamefont{Roth}},\ and\ \bibinfo
  {author} {\bibfnamefont{N.}~\bibnamefont{Paar}},\ }%
  \bibfield{journal}{%
  \Doi{10.1103/PhysRevC.75.014310}{\bibinfo {journal} {Phys. Rev. C}}\ }%
  \textbf{\bibinfo {volume} {75}},\ \bibinfo {pages} {014310} (\bibinfo {year} {2007})
  \bibAnnoteFile{NoStop}{Papakonstantinou07}%
\bibitem{Papakonstantinou10}%
  \BibitemOpen
  \bibfield{author}{%
  \bibinfo {author} {\bibfnamefont{P.}~\bibnamefont{Papakonstantinou}},
  \bibinfo {author} {\bibfnamefont{V.}~\bibnamefont{Ponomarev}}, \bibinfo
  {author} {\bibfnamefont{R.}~\bibnamefont{Roth}},\ and\ \bibinfo {author}
  {\bibfnamefont{J.}~\bibnamefont{Wambach}},\ }%
  \bibfield{journal}{%
  \bibinfo {journal} {Eur. Phys. J. A}\ }%
  \textbf{\bibinfo {volume} {47}},\ \bibinfo {pages} {1} (\bibinfo {year}
  {2011})
  \bibAnnoteFile{NoStop}{Papakonstantinou10}%
\bibitem{Papakonstantinou11}%
  \BibitemOpen
  \bibfield{author}{%
  \bibinfo {author} {\bibfnamefont{P.}~\bibnamefont{Papakonstantinou}},
  \bibinfo {author} {\bibfnamefont{V.}~\bibnamefont{Ponomarev}}, \bibinfo
  {author} {\bibfnamefont{R.}~\bibnamefont{Roth}},\ and\ \bibinfo {author}
  {\bibfnamefont{J.}~\bibnamefont{Wambach}},\ }%
  \bibfield{journal}{%
  \bibinfo {journal} {Eur. Phys. J. A}\ }%
  \textbf{\bibinfo {volume} {47}},\ \bibinfo {pages} {1} (\bibinfo {year}
  {2011})
  \bibAnnoteFile{NoStop}{Papakonstantinou11}%
\bibitem{Gunther10}%
  \BibitemOpen
  \bibfield{author}{%
  \bibinfo {author} {\bibfnamefont{A.}~\bibnamefont{G{\"u}nther}}, \bibinfo
  {author} {\bibfnamefont{R.}~\bibnamefont{Roth}}, \bibinfo {author}
  {\bibfnamefont{H.}~\bibnamefont{Hergert}},\ and\ \bibinfo {author}
  {\bibfnamefont{S.}~\bibnamefont{Reinhardt}},\ }%
  \bibfield{journal}{%
  \Doi{10.1103/PhysRevC.82.024319}{\bibinfo {journal} {Phys. Rev. C}}\ }%
  \textbf{\bibinfo {volume} {82}},\ \bibinfo {pages} {024319} (\bibinfo {year} {2010})
  \bibAnnoteFile{NoStop}{Gunther10}%
\bibitem{Gunther14}%
  \BibitemOpen
  \bibfield{author}{%
  \bibinfo {author} {\bibfnamefont{A.}~\bibnamefont{G{\"u}nther}}, \bibinfo
  {author} {\bibfnamefont{P.}~\bibnamefont{Papakonstantinou}},\ and\ \bibinfo {author}
  {\bibfnamefont{R.}~\bibnamefont{Roth}},\ }%
   \bibfield{journal}{%
  {\bibinfo {journal} {arXiv:1303.6098 [nucl-th]}}\ }%
  (\bibinfo {year} {2013})
  \bibAnnoteFile{NoStop}{Gunther14}%
\bibitem{Barbieri06}%
  \BibitemOpen
  \bibfield{author}{%
  \bibinfo {author} {\bibfnamefont{C.}~\bibnamefont{Barbieri}}, \bibinfo
  {author} {\bibfnamefont{N.}~\bibnamefont{Paar}}, \bibinfo {author}
  {\bibfnamefont{R.}~\bibnamefont{Roth}},\ and\ \bibinfo {author}
  {\bibfnamefont{P.}~\bibnamefont{Papakonstantinou}},\ }%
  \bibfield{journal}{%
  {\bibinfo {journal} {arXiv:nucl-th/0608011}}\ }%
  (\bibinfo {year} {2006})
  \bibAnnoteFile{NoStop}{Barbieri06}%
\bibitem{VanGiai81}%
  \BibitemOpen
  \bibfield{author}{%
  \bibinfo {author} {\bibfnamefont{N.~V.}\ \bibnamefont{Giai}}\ and\ \bibinfo
  {author} {\bibfnamefont{H.}~\bibnamefont{Sagawa}},\ }%
  \bibfield{journal}{%
  \Doi{10.1016/0375-9474(81)90741-7}{\bibinfo {journal} {Nucl. Phys. A}}\
  }%
  \textbf{\bibinfo {volume} {371}},\ \bibinfo {pages} {1} (\bibinfo {year}
  {1981})
  \bibAnnoteFile{NoStop}{VanGiai81}%
\bibitem{Harakeh01}%
  \BibitemOpen
  \bibfield{author}{%
  \bibinfo {author} {\bibfnamefont{M.}~\bibnamefont{Harakeh}}\ and\ \bibinfo
  {author} {\bibfnamefont{A.}~\bibnamefont{van~der Woude}},\ }%
  \emph{\bibinfo {title} {Giant Resonances}},\ \bibinfo {series} {Oxford
  Studies In Nuclear Physics}, Vol.~\bibinfo {volume} {24}\ (\bibinfo
  {publisher} {Oxford Science Publications},\ \bibinfo {year} {2001})%
  \bibAnnoteFile{NoStop}{Harakeh01}%
\bibitem{NSR1981AL05}%
  \BibitemOpen
  \bibfield{author}{%
  \bibinfo {author} {\bibfnamefont{P.~D.}\ \bibnamefont{{Allen}}}, \bibinfo
  {author} {\bibfnamefont{E.~G.}\ \bibnamefont{{Muirhead}}},\ and\ \bibinfo
  {author} {\bibfnamefont{D.~V.}\ \bibnamefont{{Webb}}},\ }%
  \bibfield{journal}{%
  \bibinfo {journal} {Nucl. Phys. A}\ }%
  \textbf{\bibinfo {volume} {357}},\ \bibinfo {pages} {171} (\bibinfo {year}
  {1981})%
  \bibAnnoteFile{NoStop}{NSR1981AL05}%
\bibitem{NSR1983PY01}%
  \BibitemOpen
  \bibfield{author}{%
  \bibinfo {author} {\bibfnamefont{R.~E.}\ \bibnamefont{{Pywell}}}, \bibinfo
  {author} {\bibfnamefont{B.~L.}\ \bibnamefont{{Berman}}}, \bibinfo {author}
  {\bibfnamefont{J.~W.}\ \bibnamefont{{Jury}}}, \bibinfo {author}
  {\bibfnamefont{J.~G.}\ \bibnamefont{{Woodworth}}}, \bibinfo {author}
  {\bibfnamefont{K.~G.}\ \bibnamefont{{McNeill}}},\ and\ \bibinfo {author}
  {\bibfnamefont{M.~N.}\ \bibnamefont{{Thompson}}},\ }%
  \bibfield{journal}{%
  \bibinfo {journal} {Phys. Rev. C}\ }%
  \textbf{\bibinfo {volume} {27}},\ \bibinfo {pages} {960} (\bibinfo {year}
  {1983})%
  \bibAnnoteFile{NoStop}{NSR1983PY01}%
\bibitem{NSR1978VA15}%
  \BibitemOpen
  \bibfield{author}{%
  \bibinfo {author} {\bibfnamefont{V.~V.}\ \bibnamefont{{Varlamov}}}, \bibinfo
  {author} {\bibfnamefont{B.~S.}\ \bibnamefont{{Ishkhanov}}}, \bibinfo {author}
  {\bibfnamefont{I.~M.}\ \bibnamefont{{Kapitonov}}}, \bibinfo {author}
  {\bibfnamefont{Z.~L.}\ \bibnamefont{{Kocharova}}},\ and\ \bibinfo {author}
  {\bibfnamefont{V.~I.}\ \bibnamefont{{Shvedunov}}},\ }%
  \bibfield{journal}{%
  \bibinfo {journal} {Yad. Fiz.}\ }%
  \textbf{\bibinfo {volume} {28}},\ \bibinfo {pages} {590} (\bibinfo {year}
  {1978})%
  \bibAnnoteFile{NoStop}{NSR1978VA15}%
\bibitem{Youngblood02}%
  \BibitemOpen
  \bibfield{author}{%
  \bibinfo {author} {\bibfnamefont{D.~H.}\ \bibnamefont{Youngblood}}, \bibinfo
  {author} {\bibfnamefont{Y.-W.}\ \bibnamefont{Lui}},\ and\ \bibinfo {author}
  {\bibfnamefont{H.~L.}\ \bibnamefont{Clark}},\ }%
  \bibfield{journal}{%
  \Doi{10.1103/PhysRevC.65.034302}{\bibinfo {journal} {Phys. Rev. C}}\ }%
  \textbf{\bibinfo {volume} {65}},\ \bibinfo {pages} {034302} (\bibinfo {year} {2002})
  \bibAnnoteFile{NoStop}{Youngblood02}%
\bibitem{Fearick14}%
  \BibitemOpen
  \bibfield{author}{%
  \bibinfo {author} {\bibfnamefont{R.~W.}\ \bibnamefont{Fearick}}, \bibinfo
  {author} {\bibfnamefont{B.}\ \bibnamefont{Erler}},\ \bibinfo
  {author} {\bibfnamefont{H.}\ \bibnamefont{Matsubara}},\ \bibinfo
  {author} {\bibfnamefont{P.}\ \bibnamefont{von Neumann-Cosel}},\ \bibinfo
  {author} {\bibfnamefont{A.}\ \bibnamefont{Richter}},\ \bibinfo
  {author} {\bibfnamefont{R.}\ \bibnamefont{Roth}},\ and\ \bibinfo {author}
  {\bibfnamefont{A.}\ \bibnamefont{Tamii}},\ }%
  \bibfield{journal}{%
  {\bibinfo {title} {in preparation}}\ }%
  (\bibinfo {year} {2014})
  \bibAnnoteFile{NoStop}{Fearick}%
\bibitem{Binder14}%
  \BibitemOpen
  \bibfield{author}{%
  \bibinfo {author} {\bibfnamefont{S.}\ \bibnamefont{{Binder}}}, \bibinfo
  {author} {\bibfnamefont{J.}\ \bibnamefont{{Lamghammer}}}, \bibinfo {author}
  {\bibfnamefont{A.}\ \bibnamefont{{Calci}}},\ and\ \bibinfo {author}
  {\bibfnamefont{R.}\ \bibnamefont{{Roth}}},\ }%
  \bibfield{journal}{%
  \bibinfo {journal} {Phys. Lett. B}\ }%
  \textbf{\bibinfo {volume} {736}},\ \bibinfo {pages} {119} (\bibinfo {year}
  {2014})%
  \bibAnnoteFile{NoStop}{Binder14}%
\bibitem{Trippel14}%
  \BibitemOpen
  \bibfield{author}{%
  \bibinfo {author} {\bibfnamefont{R.}\ \bibnamefont{Trippel}}, \bibinfo
  {author} {\bibfnamefont{P.}\ \bibnamefont{Papakonstantinou}},\ and\ \bibinfo {author}
  {\bibfnamefont{R.}\ \bibnamefont{Roth}},\ }%
  \bibfield{journal}{%
  {\bibinfo {title} {in preparation}}\ }%
  (\bibinfo {year} {2014})
  \bibAnnoteFile{NoStop}{Trippel12}%
\end{thebibliography}
\end{document}